# Security Assessment of Intel TDX with support for Live Migration

**February 2026**

*Prepared by:*

**Kirk Swidowski[1], Daniel Moghimi[2], Josh Eads[1],
Erdem Aktas[3], Jia Ma[4]**

[1] Google Cloud Vulnerability Research
[2] Google Privacy & Security Research
[3] Google Cloud Confidential Computing
[4] Google Cloud Virtualization Security

# Security Assessment of Intel TDX with Support for Live Migration

**Kirk Swidowski**[1], **Daniel Moghimi**[2], **Josh Eads**[1], **Erdem Aktas**[3], and **Jia Ma**[4]

## Summary

In the second and third quarters of 2025, Google collaborated with Intel to conduct a security assessment of Intel® Trust Domain Extensions (TDX®), extending Google's previous review and covering major changes since Intel TDX Module 1.0 – namely support for Live Migration and Trusted Domain (TD) Partitioning (nested VMs within TDs). Intel provided guidance and support, including documentation and updated TDX 1.5 source code. Unlike the previous review, this time, we had access to a compute node capable of running TDX to develop a toolkit for live testing and Proof-of-Concept (PoC) generation. Furthermore, we integrated Gemini for analysis and NotebookLM to efficiently navigate complex specifications.

This assessment resulted in the discovery of **one vulnerability that enables a VMM to fully compromise a TD**, and four vulnerabilities that enable a malicious VMM or TD to leak confidential memory of the Intel TDX Module. Several other security weaknesses and/or bugs were identified but not categorized as vulnerabilities despite having some impact on security.

Beyond presenting the technical details of multiple bugs and vulnerabilities in this report, these findings underscore that confidential computing, like other security measures, requires iterative refinement and complementary security controls to harden it, in line with a defense-in-depth approach.

Intel TDX provides the fundamental building blocks to create Trusted Execution Environments (TEEs) with a minimized Trusted Computing Base (TCB) for customers relying on a Cloud Service Provider (CSP). This means they can construct an environment with effective layers of protection, especially against insider risk. The **trustworthiness of an environment is established only after the customer attests to it** and confirms it meets their unique security policy.

While customers can delegate verification to **third-party services** (such as Intel® Tiber™ Trust Authority or Google Cloud Attestation), they must ultimately own that decision. To this end, it is the customer's responsibility to decide if a confidential computing environment is trustworthy enough for their workloads.

---

[1] Cloud Vulnerability Research

[2] Privacy & Security Research

[3] Confidential Computing

[4] Virtualization Security

*Unless otherwise stated, the vulnerabilities and bugs disclosed were present in the latest production release of the Intel TDX Module at the time of this security assessment (Q2-Q3 2025).*

*Intel has informed us that at the time of publication **all** vulnerabilities identified in this report have been remediated in versions 1.5.24/1.5.25 & 2.0.14 of the Intel TDX Module onwards (versions depend on the specific Intel platform). Intel may address other items (e.g. bugs and security weaknesses) identified in this report in subsequent releases.*

*We have verified that there has been **no evidence of active exploitation** of these vulnerabilities among Google CVM customers.*

# Introduction

Intel TDX version 1.5 supports several new features including Live Migration and TD Partitioning (nested VMs within TDs), which increase the TCB for Intel TDX. These features require new Application Programming Interfaces (APIs), workflows, and complex states, adding 34,862 lines of code to the Intel TDX module firmware compared to the 1.0 version. Within that total, 8,034 lines of code are dedicated to defining TD metadata, CPUID configurations, and the state tables required for migration. As of the time of writing, the latest Intel 1.5 specifications can be found on the [Intel Trust Domain Extensions](#) webpage.

For this project, we focused our efforts on a thorough API review (prioritizing differences since TDX 1.0) augmented by static analysis and Large Language Model (LLM) tools. Additionally, we developed a Python-based experimentation framework which was used to build a deeper understanding of complex Intel TDX flows, run experiments to test edge cases, and develop PoC exploits for discovered vulnerabilities. Similar to the previous review, we used [Frama-C](#) and [CodeQL](#) for static analysis of the code but uncovered limited findings. As LLM capabilities have significantly improved since our previous review, we used this project as an opportunity to investigate how they can assist with vulnerability discovery and variant analysis.

Previous security assessments by Google and Intel [[Intel Trust Domain Extensions (TDX) Security Review](#)] identified several vulnerabilities and weaknesses in the 1.0 version of Intel TDX. A report from Microsoft and Intel similarly focused on version 1.5 with support for Live Migration [[Technical Report of Joint Security Review By Microsoft and Intel TDX 1.5](#)], and identified several vulnerabilities in the Intel TDX module firmware.

Additionally, several research papers have been published developing side-channel attacks on Intel TDX modules [[TDXdown](#), [TDXploit](#)] and bypassing defense-in-depth mitigations aimed for countering single-stepping attack techniques.

Our work contributes to the ongoing community effort to secure the critical components in the confidential computing space and led to the discovery of several new vulnerabilities and bugs.

## Scope

In this work, we primarily focused on Intel TDX Module version 1.5 with support with Live Migration and TD Partitioning with a brief review of the Non-Persistent and Persistent SEAM Loader (NP/P-SEAMLDR). Other software components such as Intel® Software Guard Extensions (Intel® SGX) quoting enclave, and Host/Guest-related code (e.g., Linux KVM, device drivers, or specific applications) were considered outside the scope.

We did not review the MigTD as it wasn't considered ready, nor MCHECK source code because it was not available. Despite the known risk of side-channel attacks, attacks that leak memory access patterns of TDs were also outside the scope of this assessment.

## Background

The Intel TDX threat model is primarily concerned with safeguarding Trust Domains (TDs) against a malicious or compromised host environment. This includes the Virtual Machine Manager (VMM), Operating System (OS), Basic Input/Output System (BIOS), System Management Mode (SMM), legacy Virtual Machines (VMs), other TDs, and non-TD software. Despite an attacker's privileged control, **Confidentiality** and **Integrity** are maintained through a combination of architectural features, platform verification, and secure initialization.

**Availability** is not included in the security objectives because a host VMM can simply deny the Intel TDX Module and TDs the platform resources required for operation.

Key components are introduced below, with the Intel Trust Domain Extensions White Paper providing a more comprehensive description of the overall architecture.

**Secure Arbitration Mode (SEAM)**: Hosts the persistent SEAM Loader and Intel TDX Module providing protection from the host VMM, other system software, and direct-memory access (DMA) from devices using a reserved memory space identified by the SEAM Range Register (SEAMRR). New x86 Instruction Set Architecture (ISA) instructions are introduced to enter & exit, and perform SEAM operations (i.e `SEAMCALL`, `SEAMOPS`, `SEAMRET`, and `TDCALL`).

**Total Memory Encryption, Multi-Key (TME-MK)**: Provides memory encryption and integrity protection. The host VMM assigns a Host Key Identifier (HKID) and the Intel TDX Module programs a private key for that HKID into hardware, using the `PCONFIG` instruction. Only the Intel TDX Module and the TD itself are allowed to read/write associated memory.

**MCHECK**: Performs platform configuration verification (e.g., checks correct setup of SEAM Range Register and Convertible Memory Ranges); secure information storage (e.g., stores

Convertible Memory Range [CMR] table in the SEAMRR SEAMCFG region); CPU feature validation (e.g., compatible features provided by all cores and packages on a platform). This operation is undocumented, closed source, and complex. It is initiated as part of a μCode patch by writing the `IA32_BIOS_UPDT_TRIG` Model Specific Register (MSR).

**SEAM Loader (SEAMLDR)**: The NP-SEAMLDR is an Authenticated Code Module (ACM) responsible for verification and loading of the P-SEAMLDR. The P-SEAMLDR runs in a subrange of the SEAMRR and is responsible for verification and loading or updating of the Intel TDX Module.

**Intel TDX Module**: Software that runs in SEAM and used by a VMM to support TD operations. This runtime software is responsible for providing TD security.

**TD Attestation Software**: TD software can request a local attestation report (TDReport) which provides information on platform configuration and software measurements that can be used to establish trust of the execution environment. An Intel SGX enclave (TDQuoting Enclave) is used to sign TDReports and generate a remote attestation report (TDQuote). These reports are also used for Live Migration to establish trust between source and destination Intel TDX Modules. Intel provides general-certification infrastructure to verify that a TDQuote was generated by a genuine Intel platform, but customers are expected to verify the content of the attestation report to establish trust based on their security policy.

With the introduction of Live Migration, the Intel TDX Module is now required to maintain Confidentiality and Integrity of TDs not just on the platform running the TD, but between multiple platforms, and during the migration process. Private memory, non-memory state, and control state of a TD **_must not_** be disclosed or modified by untrusted software, which is used extensively to facilitate TD movement. Additionally, only a single instance of the TD being migrated is allowed to run at any point in time.

Components inside the Intel TDX TCB (e.g., Intel TDX Module, SEAM Loaders, and CPU Hardware) are each assigned a Security Version Number (SVN). These are included in attestation and verified at startup to be greater than or equal to some threshold value.

In the case where vulnerabilities are discovered in TCB components, they are fixed and the associated SVN is updated. Depending on the severity of the issue and other factors, a process known as TCB-Recovery (TCB-R) can be performed to ensure that older versions of components reflect their reduced security level via an attestation process. Intel's [Trusted Computing Base Recovery of Intel Trusted Execution Environments](#) web page describes the process in detail.

Next, we describe components of the Intel TDX Module that are the most relevant to our work.

## State Machines

The Intel TDX Module uses two primary state machines for a TD: *lifecycle* and *operation*.

**Lifecycle State**: Tracked in the Trust Domain Root (TDR) management fields as lifecycle_state and has the following states: `TD_HKID_ASSIGNED`, `TD_KEYS_CONFIGURED`, `TD_BLOCKED`, and `TD_TEARDOWN`. Most of a TD's lifetime is spent in the `TD_KEYS_CONFIGURED` state. `TD_HKID_ASSIGNED` is the initial state when the TD is first created and `TD_BLOCKED`/`TD_TEARDOWN` are used during resource reclamation.

**Operation State**: Tracked in the Trust Domain Control Structure (TDCS) management fields as `op_state`. This state machine is more complex and is primarily checked with `check_state_map_tdcs_and_lock` throughout the codebase. A full breakdown of the APIs allowed in each op_state is provided in [Appendix A](#). This state machine is largely used to restrict API access unless a TD is in an acceptable `op_state`.

The Secure Extended Page Tables (SEPT) entries also include a state to ensure that a given operation is only allowed if an entry is in a given state. This is usually checked with the `sept_state_is_seamcall_leaf_allowed` function.

## TD Migration

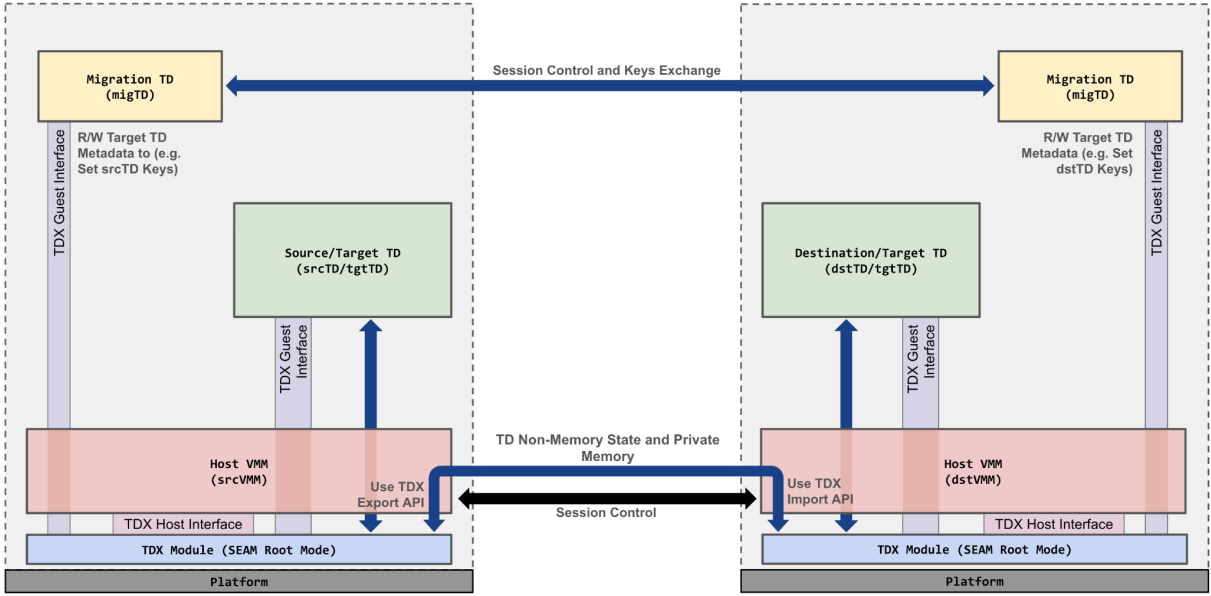

**Figure 1: TD Migration Components and Process**
(Adapted from Figure 2.2 in the [Intel TDX Module Architecture Specification: TD Migration](#))

TD migration enables the secure relocation of an executing TD between Intel TDX platforms in an untrusted environment. CSPs utilize this feature to relocate TDs and meet customer

Service Level Agreements (SLAs) while maintaining the ability to perform critical maintenance tasks (e.g., upgrade software, replace hardware, and patch firmware).

The figure above shows the components and processes of a migration. Regardless of the migration type: **cold** (i.e. TD is suspended, transferred, and resumed) or **live** (i.e. TD remains running with a brief blackout period) a similar process is performed with live migration having a few distinct phases for memory transfer: In-Order Memory, Blackout, and Out-Of-Order Memory.

**In-Order Memory Migration Phase**: Occurs while the Source TD (srcTD) may continue to run and modify its memory and non-memory state. Order is critical in this phase to ensure that exports of the same memory are imported in the correct order (i.e. oldest first and newest last).

**Blackout Period**: The srcTD is stopped and the mutable non-memory state (e.g., TD, and VCPUs) are transferred.

**Out-of-Order Memory Migration Phase**: Occurs after the Blackout Period and the Destination TD (dstTD) can start executing. Memory can be transferred in any order and allows on-demand (e.g., EPT violation prioritization) based memory transfers.

The host VMM is responsible for managing resources for the TDs and during a migration interacting with the Intel TDX Module to export and import migration bundles. Migration bundles are private memory or non-memory state for a TD, and are both encrypted and integrity protected. Before migration bundles can be sent, the Destination VMM (dstVMM) creates a template TD to import the migration bundles into. This is accomplished with multiple calls to the Intel TDX Module via the Intel TDX Host Interface (i.e. using the `SEAMCALL` instruction) to assign an HKID, assign memory, and initialize data structures.

Before the Target TD (tgtTD) can be migrated, a Migration TD (migTD) is created by both the Source VMM (srcVMM) and dstVMM. This TD plays a critical role in enabling secure and verifiable relocation of the tgtTD by examining the Intel TDX Module's capabilities and attestation evidence. This information is checked on both platforms against the migration policy and includes: acceptable TD attributes, allowed SVNs, and supported migration protocol version. The migTDs each communicate a Migration Session Key (MSK), which are AES-GCM-256 keys generated by the Intel TDX Module, to the other. When the destination migTD receives the MSK it sets it in the dstTD via the Intel TDX Guest Interface (i.e. using the `TDCALL` instruction).

The tgtTD is the TD being migrated, the srcTD is running pre-migration and the dstTD is running post-migration. The migTD is bound to the srcTD before TD measurements are finalized during TD initialization. The tgtTD doesn't perform any special activities during a migration and is unaware of the process taking place.

This process is similar to but considerably more complex than the migration of a legacy VM. With a legacy migration the host VMM is included in the VM's TCB and has full access to memory and non-memory state. Because of this the srcVMM and dstVMM can perform the migration without additional software (i.e. the migTD and Intel TDX Module). Instead the host VMMs perform the necessary verification and authentication and provide a secure channel to communicate migration activities.

*Note: The purpose of TD migration is to enable relocation of a TD across physical platforms but technically these activities could take place on a single platform.*

## Creation Workflows

TD migration introduces an additional path for constructing TDs where import API are used (e.g., `tdh_import_state_immutable`, `tdh_import_state_td`, and `tdh_import_state_vp`) instead of the initialization API (e.g., `tdh_mng_init` and `tdh_vp_init`).

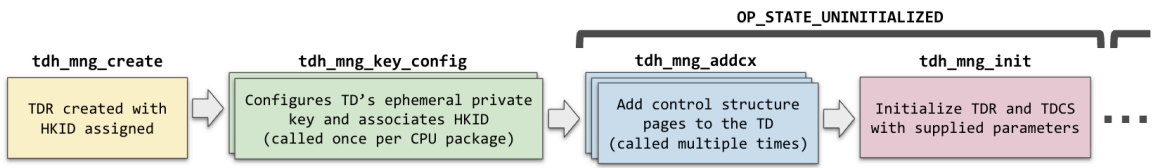

**Figure 2: Uninitialized Build Sequence**

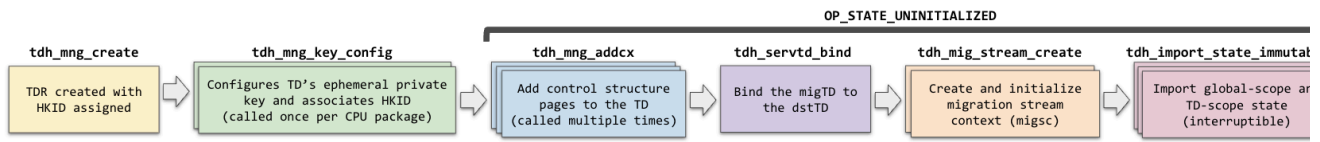

**Figure 3: Uninitialized Import Sequence**

(Adapted from Figure 2.4 in the Intel TDX Module Architecture Specification: TD Migration)

The traditional Build Sequence, shown in Figure 2, and the Import Sequence, shown in Figure 3, start the same way with a call to `tdh_mng_create` with a Host Physical Address (HPA) for the TDR and HKID supplied as parameters. The TDR page is initialized and the HKID is set to `KOT_STATE_HKID_ASSIGNED`. `tdh_mng_key_config` is used to program the HKID and encryption key into the TME-MK. `tdh_mng_addcx` is then called multiple times to add pages and initialize them to be used for the TD Control Structures (TDCS).

## Build Sequence

The Build Sequence at this point diverges from the common flow by calling `tdh_mng_init` to initialize the global-scope of the TD and TD-scope state shared by the Virtual Processors (VPs) (e.g., L2 Virtual Machine [VM] count, owner measurement fields, CPUID configuration, Virtual Machine eXtensions [VMX] controls, and MSR bitmap). Data to populate the TD originates from `td_module_global_t` and `td_module_local_t` structures located inside the Intel TDX Module as well as a td_params_t structure supplied by the host VMM. Upon successful completion of that call the `op_state` of the TD is set to `OP_STATE_INITIALIZED`.

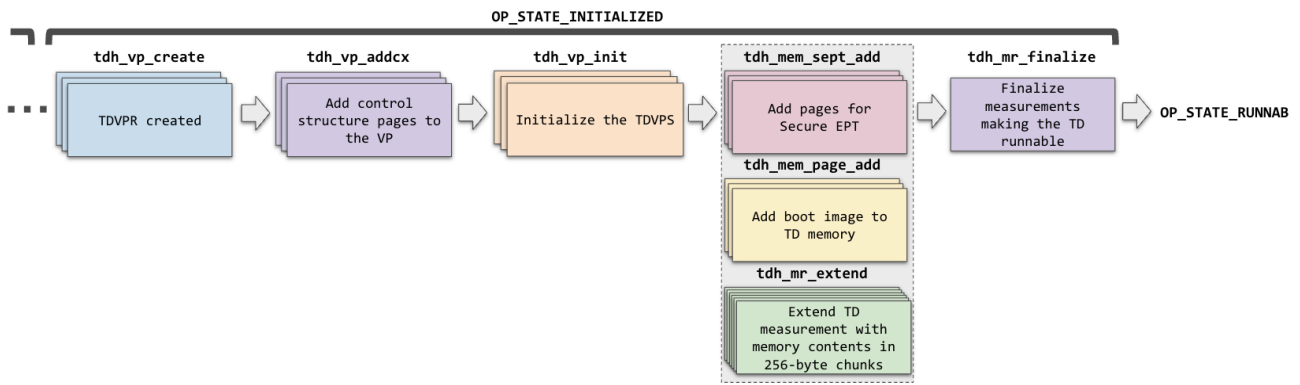

**Figure 4: Initialized Build Sequence**

From the `OP_STATE_INITIALIZED` state the host VMM can proceed to create and initialize VPs using a sequence of calls to `tdh_vp_create`, `tdh_vp_addcx`, and `tdh_vp_init`. Private memory is added with a series of calls to tdh_mem_sept_add to construct the paging hierarchy and `tdh_mem_sept_add` to add a memory page and populate the contents. `tdh_mr_extend` is used to extend the TD measurement with the contents of the added pages. Adding VPs and memory can be done in any order and does not need to be performed by the same LP. Completion of TD initialization occurs when `tdh_mr_finalize` is called which sets the `op_state` to `OP_STATE_RUNNABLE`.

## Import Sequence

The Import Sequence is more complex requiring a migTD to be bound to the tgtTD using `tdh_sertd_bind` or `tdh_servtd_prebind` and creating Migration Stream Contexts (MigSCs) with `tdh_mig_stream_create`. With that completed, the migTDs on the source and destination exchange MSKs as described in the [TD Migration](#) section. After the exchange is completed the dstVMM receives encrypted migration bundles to initialize the dstTD by calling `tdh_import_state_immutable`. This API performs similar initialization to `tdh_mng_init` but using encrypted migration bundles passed by the dstVMM. As this can be complex and time-consuming (e.g., decrypting the migration bundle, importing metadata lists, and processing additional authenticated data) the Intel TDX Module checks for pending interrupts during operation. When a pending interrupt is present the `migsc_t` structure is used to preserve state and return execution to the host VMM. The host VMM can process interrupts, perform other activities, and resume the import at a later time by calling `TDH_IMPORT_STATE_IMMUTABLE`. When the import activity is resumed the `migsc_t` structure is used to restore state and check that other Intel TDX Module import activities have not taken place. Once the immutable state is completely imported the `op_state` of the TD is set to `OP_STATE_MEMORY_IMPORT`.

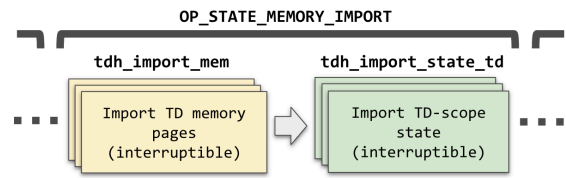

**Figure 5: Memory Import Sequence**
(Adapted from Figure 2.4 in the Intel TDX Module Architecture Specification: TD Migration)

This `op_state` starts what is known as the "In-Order Memory Import Phase". Secure EPT structures are created and memory is imported using `tdh_import_mem`. Each page imported is decrypted and copied into the TD. Page attributes and the location are verified to ensure they are also the same between source and destination. Up to this point the srcTD is in the live export state, allowed to run, and change TD memory and non-memory state. Mutable TD-scope state can only be exported during the Intel TDX-imposed Blackout which is started when `tdh_export_pause` is called on the srcTD which prevents the srcTD from running. `tdh_import_state_td` can then be used to import the mutable TD state and change the `op_state` to `OP_STATE_STATE_IMPORT` upon completion. Both APIs are considered time-consuming and support interruption similar to `tdh_import_state_immutable`. Shared memory is migrated using traditional host VMM workflows.

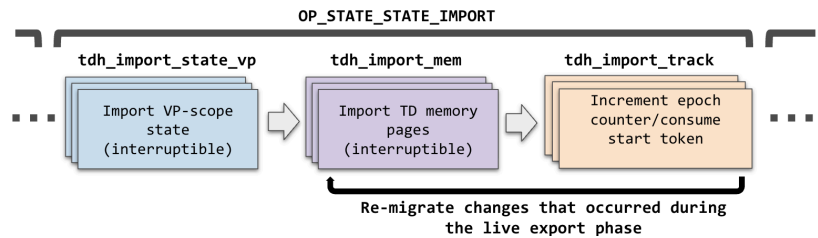

**Figure 6: State Import Sequence**
(Adapted from Figure 2.4 and 2.5 in the Intel TDX Module Architecture Specification: TD Migration)

During `OP_STATE_STATE_IMPORT` VP-scope state is imported using `tdh_import_state_vp` and memory can continue to be imported using `tdh_import_mem`. This `op_state` can continue to create Secure EPT structures as necessary to perform the memory import activities. With the TD in the Intel TDX-imposed Blackout, memory marked dirty during the source TD's live export phase is re-migrated. `tdh_import_track` consumes a migration epoch token created by `tdh_export_track` on the source and either starts a new epoch leaving the TD in the current state or changes to the `OP_STATE_POST_IMPORT op_state` when the start token is received.

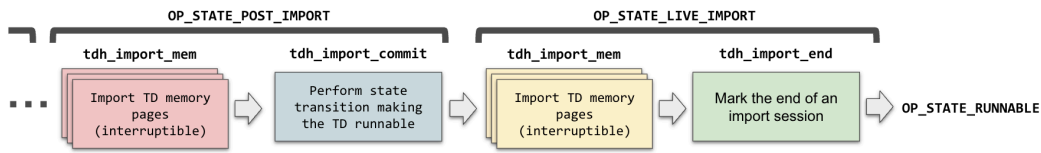

**Figure 7: State Import Sequence**
(Adapted from Figure 2.5 in the Intel TDX Module Architecture Specification: TD Migration)

The `OP_STATE_POST_IMPORT op_state` starts the "Out-of-Order Memory Import Phase" and is used to support post-copy migration. At this point the srcTD is no longer runnable and memory no longer needs to be tracked for freshness. Memory can continue to be imported and when `tdh_import_commit` completes the Intel TDX-Imposed Blackout is over. This means that the dstTD can be run. Post-copy migration allows the dstTD to start executing before all memory pages have been transferred. After `tdh_import_commit` the `op_state` is changed to `OP_STATE_LIVE_IMPORT` and memory can continue to be imported. If the TD attempts to use non-present memory an EPT violation occurs and the dstVMM can decide how to prioritize the remaining pages to be migrated. The "Out-of-Order Memory Import Phase" ends along with the migration when `tdh_import_end` is called. This changes the TD `op_state` to `OP_STATE_RUNNABLE`.

## Migration Bundles

During a migration, bundles are created by the Intel TDX Module through the export API (i.e. `tdh_export_state_immutable`, `tdh_export_state_td`, `tdh_export_state_vp`, and `tdh_export_mem`) and provided to the host VMM. These bundles include Migration Bundle Metadata (MBMD) and migration data. The MBMD is integrity protected and the migration data is encrypted to preserve confidentiality. The migration data consists of one or more migration pages and each page is represented as a Metadata List.

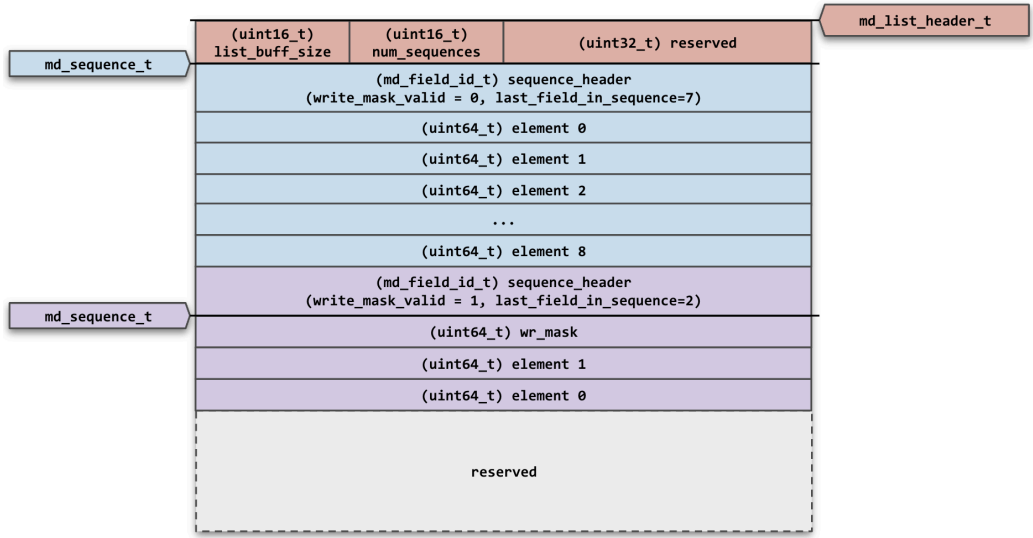

**Figure 8: Non-Memory State Migration List**
(Adapted from Figure 7.1 in the Intel TDX Module Architecture Specification: TD Migration)

The `md_list_header_t` includes the `list_buff_size` as a `uint16_t` and `num_sequences` as a `uint16_t`. `list_buff_size` is the total number of bytes for the `md_list_t` and includes the size of `md_list_header_t`. The Intel TDX Module constrains the size of the `md_list_t` at runtime to be less than or equal to `_4KB`. The `md_list_header_t` and `md_list_t`, as defined by the Intel TDX Module, are provided below.

```
typedef union md_list_header_u {
  struct {
    uint16_t list_buff_size;
    uint16_t num_sequences;
    uint32_t reserved;
  };
  uint64_t raw;
} md_list_header_t;

typedef union md_list_u {
  struct {
    md_list_header_t hdr;
    uint8_t body[_4KB - sizeof(md_list_header_t)];
  };
  uint8_t raw[_4KB];
} md_list_t;
```

After the header is an array of `md_sequence_t` structures, located in the `body` of the

`md_list_t` above, each starting with a `sequence_header`. The `sequence_header` is of type `md_field_id_t`, and the figure below provides a simplified layout.

```c
typedef union md_field_id_u {
...
  struct {
    uint32_t field_code : 24;    // Bits 0:23
    uint32_t reserved_0 : 8;      // Bits 24:31
  };
...
  struct {
    uint32_t element_size_code     : 2;    // Bits 33:32
    uint32_t last_element_in_field : 4;    // Bits 37:34
    uint32_t last_field_in_sequence : 9;   // Bits 46:38
    uint32_t reserved_1            : 3;    // Bits 49:47
    uint32_t inc_size             : 1;    // Bit 50
    uint32_t write_mask_valid      : 1;    // Bit 51
    uint32_t context_code          : 3;    // Bits 54:52
    uint32_t reserved_2           : 1;    // Bit 55
    uint32_t class_code            : 6;    // Bits 61:56
    uint32_t reserved_3           : 1;    // Bit 62
    uint32_t ignored              : 1;    // Bit 63
  };
...
} md_field_id_t;
```

Important fields are in bold with `field_code`, `context_code`, and `class_code` being used to uniquely identify an entry to be imported. The `last_field_in_sequence` and `write_mask_valid` fields are used during import to determine how many entries are represented in a sequence and if the first element in the sequence should be used as a `write_mask` when importing values.

## Identified Vulnerabilities

The following table highlights the findings that were confirmed as vulnerabilities. Out of the 5 reported vulnerabilities, we discovered one high severity vulnerability that enables a VMM to fully compromise a TD, and four vulnerabilities that enable a malicious VMM or TD to leak confidential memory of the Intel TDX Module. We also found several other security weaknesses that were not attributed to Common Vulnerability and Exposures (CVE) identifiers despite some impact on security, which are discussed in the next section.

| Identifier | Type | Score | Description |
|---|---|---|---|
| CVE-2025-30513 | Time-of-Check / Time-of-Use | 7.9 | Migratable TD can become debuggable during migration |
| CVE-2025-32007 | Out-of-bounds Read | 4.4 | Metadata sequence parsing leads to an integer underflow |
| CVE-2025-27572 | Speculative Out-of-bounds Read | 4.1 | Speculative out-of-bounds read in guest `RDMSR` and `WRMSR` handlers |
| CVE-2025-32467 | Speculative Out-of-bounds Read | 4.1 | Speculative out-of-bounds read in host HKID free and VP flush API |
| CVE-2025-27940 | Speculative Out-of-bounds Read | 4.1 | Speculative out-of-bounds read in host API to prebind and bind a service TD |

**Intel provided an updated Intel TDX Module with all identified vulnerabilities addressed to customers as part of Intel Platform Update (IPU) 2026.1 or other product sustaining releases between September and December 2025.**

After the planned February 10, 2026 public disclosure of these issues, Intel will initiate a TCB Recovery process that enables relying parties to verify whether the latest updates have been deployed on the platform they are using, make security decisions, and establish or re-establish trust with the platform.

## Vulnerability 1: Migratable TD can Become Debuggable During Migration

**Intel Technical Advisory for CVE-2025-30513**: In certain Intel TDX modules, the TDH.MNG.INIT API may be executed after the TDH.IMPORT.STATE.IMMUTABLE state is initiated, potentially enabling an exploit in which a migratable TD is imported as a debuggable TD.

**Attack Scenario**: A compromised dstVMM participates in the migration of a tgtTD. The tgtTD is correctly configured by an uncompromised srcVMM, had attestation verified, and was provided with confidential user data. The TD is migratable and a trusted migTD has been bound to it, the TD is **not** debuggable.

*In this situation a dstVMM can exploit a Time-of-Check to Time-of-Use vulnerability to change the TD's attributes from `migratable` to `debug` as the TD's immutable state is being imported!*

**Exploitation of this attack breaks confidentiality and integrity of a migratable TD.** Once the TD is marked debuggable the host VMM is given complete access to the TD's private memory and non-memory state.

Interrupting a `tdh_import_state_immutable` operation and interleaving a call to `tdh_mng_init` allows a TD's immutable state, including the `attributes`, to be modified after being imported. The import operation can then be resumed with the Intel TDX Module unaware of any modification. The rest of this section details the root cause of the vulnerability and provides a proof-of-concept exploit.

The `attributes` field within a TD is a critical 64-bit bitmap that is used to specify various characteristics and is included in attestation. Three groupings defined:
- **TD Under Debug (TUD):** *Any bit set in this group renders the TD untrusted*. Only the `debug` flag is defined and when set provides the host VMM access to TD state, VCPU state and private memory.
- **Security (SEC):** Bits in this group impact TD security and include the `migratable` flag.
- **Other (OTHER):** Bits in this group do not impact TD security. The `perfmon` flag is defined and allows the TD to use performance monitoring capabilities.

The [Creation Workflows](#) section describes the similarities and differences between building and importing a TD. In either situation the TD has an `op_state` of `OP_STATE_UNINITIALIZED` allowing `tdh_mng_init` or `tdh_import_state_immutable` to be called.

```
// tdh_mng_init.c
check_lock_and_map_explicit_tdr(tdr_pa, OPERAND_ID_RCX, TDX_RANGE_RW,
TDX_LOCK_EXCLUSIVE, PT_TDR, &tdr_pamt_block, &tdr_pamt_entry_ptr, &tdr_locked_flag,
&tdr_ptr);
...
check_state_map_tdcs_and_lock(tdr_ptr, TDX_RANGE_RW, TDX_LOCK_NO_LOCK, false,
TDH_MNG_INIT_LEAF, &tdcs_ptr);
```

```
// tdh_import_state_immutable.c
check_lock_and_map_explicit_tdr(tdr_pa, OPERAND_ID_RCX, TDX_RANGE_RW,
TDX_LOCK_EXCLUSIVE, PT_TDR, &tdr_pamt_block, &tdr_pamt_entry_ptr, &tdr_locked_flag,
&tdr_p};
...
check_state_map_tdcs_and_lock(tdr_p, TDX_RANGE_RW, TDX_LOCK_NO_LOCK, false,
TDH_IMPORT_STATE_IMMUTABLE_LEAF, &tdcs_p);
```

Each API uses `check_lock_and_map_explicit_tdr` to take an exclusive lock on the TDR and map it into the TDX module's linear address space. Next, they call `check_state_map_tdcs_and_lock` to ensure the `op_state` is valid for the called API.

`tdh_import_state_immutable` breaks the supplied migration bundle up into multiple `md_list_t` structures and iterates over each `md_list` calling `md_write_list` to write entries into the TDR and TDCS. When `md_write_list` returns it checks if pending host interrupts are present using `is_interrupt_pending_host_side`. If `true` the state is saved into `migsc_p` and execution is returned to the host VMM with a status code of `TDX_INTERRUPTED_RESUMABLE`. The `op_state` is **not** modified when an interruption occurs.

```
api_error_type tdh_import_state_immutable(uint64_t target_tdr_pa, uint64_t
hpa_and_size_pa, uint64_t page_or_list_pa, uint64_t migs_i_and_cmd_pa) {
  ...
  page_list_info.raw = page_or_list_pa;
  ...
  do {
    md_list_pa.raw = page_list_p[page_list_i].raw;
    ...
    md_list_hdr_p = (md_list_header_t*)map_pa(md_list_pa.raw_void, TDX_RANGE_RO);
    if (aes_gcm_decrypt(&migsc_p->aes_gcm_context, (uint8_t*)md_list_hdr_p,
(uint8_t*)&md_list, _4KB) != AES_GCM_NO_ERROR) {
      fatal_error(FATAL_ERROR_ID_150, FATAL_INFO_FORMAT_BASIC_INFO, NULL);
    }
    ...
    if (!sys_imported) {
      ...
      return_val = md_write_list(MD_CTX_SYS, field_id, ...);
      ...
    }
    else { // Import the TD metadata list:
      return_val = md_write_list(MD_CTX_TD, field_id, ...);
    }
    ...
    if ((page_list_i <= page_list_info.last_entry) &&
is_interrupt_pending_host_side()) {
      ...
      return_val = TDX_INTERRUPTED_RESUMABLE;
      goto EXIT;
    }
  } while ((uint64_t)page_list_i <= page_list_info.last_entry);
  ...
  tdcs_p->management_fields.op_state = OP_STATE_MEMORY_IMPORT;
  return_val = TDX_SUCCESS;
EXIT:
  ...
  return return_val;
```

At the time of writing, the immutable state metadata bundle is 12KB (composed of three `md_list_t` structures). A simplified version is shown in Figure 5.

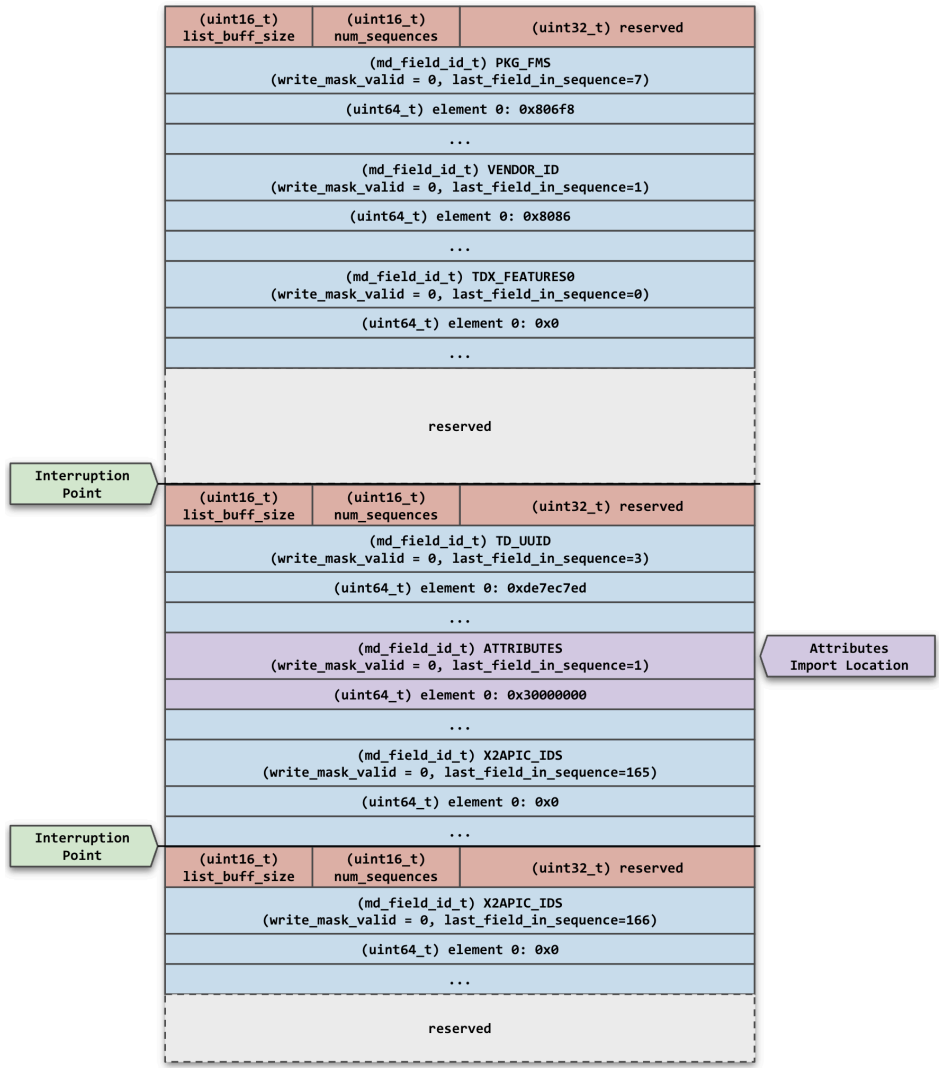

**Figure 9: Simplified Immutable State Metadata Bundle**

The first `md_list` contains fields associated with the TDX module (e.g., `BUILD_NUM`, `VENDOR_ID`, and `TDX_FEATURES0`) and does not consume an entire 4KB page (remaining space is filled with zeros). The second and third `md_list` structures contain fields associated with the TD (e.g., `TD_UUID`, `NUM_VCPUS`, `ATTRIBUTES`, and `X2APIC_IDS`). This provides two interruption points with everything except the `X2APIC_IDS` describing the TD being located in the second `md_list`.

The TD attributes are imported with an `import_mask` of `(-1ULL & 0xFFFFFFFFFFFFFFFFULL)` but has `.special_wr_handling` set to `true` as shown below. This means that any bit in the 64bit field can be set but additional processing occurs before the value is actually written into the TD.

```
// ATTRIBUTES // 15
.field_id = { .raw = 0x1110000300000000 },
...
.export_mask = (-1ULL & 0xFFFFFFFFFFFFFFFFULL), .import_mask = (-1ULL &
0xFFFFFFFFFFFFFFFFULL),
.special_rd_handling = false, .special_wr_handling = true,
.mig_export = MIG_MB, .mig_import = MIG_MB
```

The special write handling that takes place in `md_td_write_field` is used to call `verify_td_attributes`. If this function returns `false`, the write does not occur and instead returns `TDX_METADATA_FIELD_NOT_VALID`.

```
api_error_code_e md_td_write_field(md_field_id_t field_id, const md_lookup_t*
entry, md_access_t access_type,
  ...
      case MD_TDCS_ATTRIBUTES_FIELD_ID:
         ...
         td_param_attributes_t attributes;
         attributes.raw = value[0] & combined_wr_mask;
         if (!verify_td_attributes(attributes, is_import)) {
           return TDX_METADATA_FIELD_VALUE_NOT_VALID;
         }
         break;
  ...
```

`verify_td_attributes` when called by the write handler and has `is_import` set to `true` which ensures that the `migratable` flag is set and that `debug` and `perfmon` flags are clear.

```
bool_t verify_td_attributes(td_param_attributes_t attributes, bool_t is_import)
  ...
  if (attributes.migratable) {
    // A migratable TD can't be a debug TD and does not support PERFMON
    if (attributes.debug || attributes.perfmon) { return false; }
  }
  else if (is_import){
      // TD must be migratable on import flow
      return false;
  }
  ...
  return true;
```

```
}
```

Looking at the other initialization path, `tdh_mng_init` uses the host VMM supplied `td_params_t` structure called `td_params_ptr` to initialize a TD. `read_and_set_td_configurations` is used to validate and write parameters into it. If that function returns a value other than `TDX_SUCCESS` then `goto EXIT` is executed and returns the error code to the host VMM. The `op_state` is **_not_** modified when an error occurs.

```
api_error_type tdh_mng_init(uint64_t target_tdr_pa, uint64_t target_td_params_pa,
uint64_t event_filters_info_params) {
  ...
  td_params_pa.raw = target_td_params_pa;
  ...
  td_params_ptr = (td_params_t *)map_pa((void*)td_params_pa.raw, TDX_RANGE_RO);
  ...
  return_val = read_and_set_td_configurations(tdr_ptr, tdcs_ptr, td_params_ptr);
  if (return_val != TDX_SUCCESS) {
    TDX_ERROR("read_and_set_td_configurations failed\n");
    goto EXIT;
  }
  ...
  tdcs_ptr->management_fields.op_state = OP_STATE_INITIALIZED;
EXIT:
  ...
  return return_val;
```

The first parameter that `read_and_set_td_configuration` checks is the `attributes`. It uses `verify_td_attributes` with `is_import` set to `false`. This allows a TD to be configured with the `debug` and/or `perfmon` flags set as long as the `migratable` flag is clear. If `verify_td_attributes` returns `true` the `tmp_attributes` are assigned to the TD.

```
static api_error_type read_and_set_td_configurations(tdr_t * tdr_ptr,
  ...
  tmp_attributes.raw = td_params_ptr->attributes.raw;
  if (!verify_td_attributes(tmp_attributes, false)) {
    return_val = api_error_with_operand_id(TDX_OPERAND_INVALID,
OPERAND_ID_ATTRIBUTES);
    goto EXIT;
  }
  tdcs_ptr->executions_ctl_fields.attributes.raw = tmp_attributes.raw;

  tdcs_ptr->executions_ctl_fields.td_ctls.pending_ve_disable =
tmp_attributes.sept_ve_disable;
```

```
tmp_xfam.raw = td_params_ptr->xfam;
if (!check_xfam(tmp_xfam)) {
  return_val = api_error_with_operand_id(TDX_OPERAND_INVALID, OPERAND_ID_XFAM);
  goto EXIT;
}
...
```

If a later check, for example `check_xfam`, fails the TD's `attributes` have already been modified, are not restored, and the `op_state` is left as `OP_STATE_UNINITIALIZED`.

To summarize the root causes of this vulnerability that lead to exploitation:

1. `tdh_mng_init` does not modify the `op_state` on failure but **does** modify TD state
2. `tdh_import_state_immutable` is interruptible but does **not** correctly validate imported state after the entire migration bundle is imported (relying instead on `md_td_write_field` checks during the import of each field)
3. `tdh_import_state_immutable` only modifies the `op_state` after the entire migration bundle is imported
4. `migsc_t` is used to track migration state in the migration API but other APIs are unaware of it

A compromised dstVMM can exploit these conditions to convert a migratable TD to a debuggable TD by performing the following steps:

1. Create an interrupt storm targeting the LP that will perform the immutable state import
2. Call `tdh_import_state_immutable` with the immutable state and MBMD
   a. Goto step 3 if the Intel TDX module returns `TDX_INTERRUPTED_RESUMABLE`
   b. Goto step 4 if it completes with `TDX_SUCCESS` and step 3 was previously executed
3. Call `tdh_mng_init` with `attributes.debug` set and an invalid `xfam` value
   a. Intel TDX module returns with `TDX_OPERAND_INVALID` - `OPERAND_ID_XFAM`
   b. Goto step 2
4. Call `tdh_mng_rd` to read the `MIG_DEC_WORKING_KEY`

At this point the migration can be continued but with the dstVMM having complete access to the dstTD. The dstVMM can use the `MIG_DEC_WORKING_KEY`, which is only allowed to be read by the dstVMM when the TD is debuggable, to decrypt migration bundles received by the srcVMM and modify them before performing additional import operations.

As the migration bundles represent the entire state of the TD (e.g., non-memory state and private memory) and because the dstVMM can decrypt them, the TD's confidentiality is compromised.

One side effect of calling `tdh_mng_init` is that `num_vcpus` is initialized to `0` before the `attributes` are set which prevents VP import. The dstTD can still transition through expected operation states without issue because the `tdh_import_track` API only verifies `num_migrated_vcpus` and `num_vcpus` are equal.

Even with that side effect Integrity is compromised because the migration can be completed with the dstTD in a different state than the srcTD. Furthermore, an attacker can proceed to impact integrity by:

1. Using the migration bundles to construct another TD and destroy the dstTD once completed
2. Abort the migration and use the confidential data to attack the srcTD (e.g., extract credentials, persistent storage keys, or encrypted connection details).

***A simplified attack scenario where the srcVMM and dstVMM are the same platform is also possible.*** In this case the complexity of exploitation is reduced and provides an additional live monitoring capability. In this case the host VMM initiates a migration but does not complete it. Instead, it stays in the `OP_STATE_LIVE_EXPORT` op_state, using `tdh_export_mem` to access private memory as needed.

The PoC, shown in Figures 10 and 11, performs the attack described above to extract the `MIG_DEC_KEY`, using the [TDXplore Toolkit](TDXplore Toolkit). A single host VMM and migTD was used as both the source and destination to simplify demonstration.

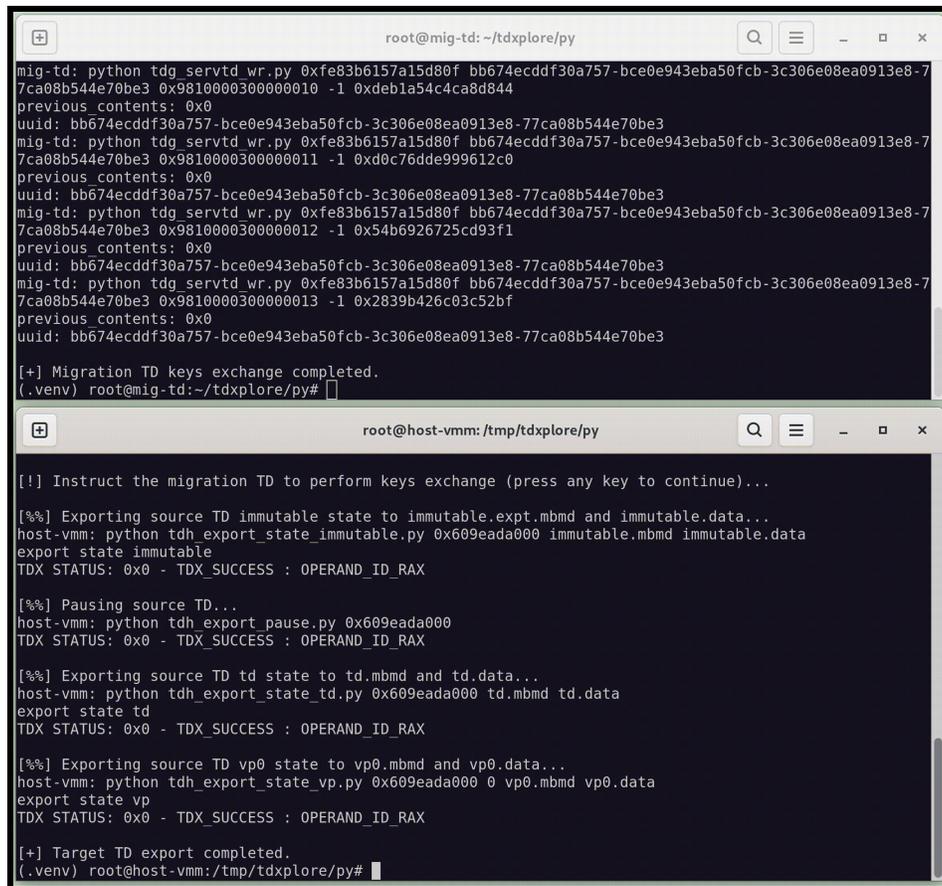

**Figure 10: Uncompromised Setup and Export Operations**
*Note: An animated version of the figure is provided here.*

Figure 10 includes a terminal for the migTD (top) and host VMM (bottom). The host VMM is **not** part of the TCB and only used to perform expected resource management activities. The migTD is bound to both the srcTD and dstTD and completes the required keys exchange within its context.

**Figure 11: Exploited Import Operations Making the TD Debuggable**
*Note: An animated version of the figure is provided here.*

Figure 11 contains a terminal on the host VMM and is used to run the exploit. Once the exploit completes it proceeds to print the `attributes` showing the imported TD has the `debug` flag set and prints the `MIG_DEC_KEY`, using the `tdh_mng_rd` API. The `MIG_DEC_KEY` field is only accessible to the host VMM when the TD is marked debuggable.

**Remediation**: This vulnerability was fixed by introducing a new `op_state`, called `START_IMPORT`, to prevent the non-import path from being taken after an immutable import has been started.

This vulnerability is only exploitable when a Trust Domain (TD) is configured to be migratable. Until this fix is fully deployed, customers **should check their attestation report** to verify CVMs are built with migration support disabled. Even after the fix is deployed, customers should continue to carefully review their migration policy and environment configuration. This is to ensure they meet the minimum Security Version Number (SVN) and the specific confidentiality requirements for their use case and threat model. A misconfigured migration policy could allow a CVM to be moved to a host with a vulnerable environment, exposing it to known threats.

## Vulnerability 2: Metadata Sequence Parsing Leads to an Integer Underflow

**Intel Technical Advisory for CVE-2025-32007**: In certain scenarios, a VMM may generate a migration stream that interacts with the Intel TDX Module in a way that could lead to memory access beyond allocated boundaries. This condition may result in unintended reads of system memory, potentially revealing privileged information.

**Attack Scenario**: A compromised host VMM creates a template TD to import migration bundles into. A malicious migTD is created and bound to the template TD. The MSK is provided by the migTD to the host VMM so it can construct and encrypt migration bundles for import.

*In this situation, the host VMM can exploit an integer underflow condition that occurs during import, allowing up to 8KB of Out-Of-Bounds (OOB) data to be read from the current LP's stack in the Intel TDX Module.*

**Exploitation of this vulnerability allows OOB reads of data in the Intel TDX Module.** With this an attacker is able to: bypass Address Space Layout Randomization (ASLR), leak the global stack canary, read the contents of the LP's shadow stack, and read data from an adjacent LP's stack in the Intel TDX Module.

There are three APIs used to import non-memory TD state in the Intel TDX Module: `tdh_import_state_immutable`, `tdh_import_state_td`, and `tdh_import_state_vp`. These APIs share a similar flow and the vulnerability is reachable by any of them. `tdh_import_state_vp` will be used to describe how to reach the root cause of the vulnerability and demonstrate a PoC exploit.

```
api_error_type tdh_import_state_vp(uint64_t target_tdvpr_pa, ...uint64_t
hpa_and_size_pa, uint64_t page_or_list_pa, uint64_t migs_i_and_cmd_pa) {
  ...
  md_list_header_t *md_list_hdr_p = NULL;
  ...
  md_list_t md_list;
  ...
  page_list_pa.raw = 0;
  page_list_pa.page_4k_num = page_list_info.hpa;
  page_list_p = (pa_t *)map_pa(page_list_pa.raw_void, TDX_RANGE_RO);
  ...
  do {
    ...
    md_list_pa.raw = page_list_p[page_list_i].raw;
    md_list_hdr_p = (md_list_header_t *)map_pa(md_list_pa.raw_void, TDX_RANGE_RO);

    if (aes_gcm_decrypt(&migsc_p->aes_gcm_context, (uint8_t*)md_list_hdr_p,
(uint8_t*)&md_list, _4KB) != AES_GCM_NO_ERROR) {
```

```
    ...
    if (md_list.hdr.list_buff_size > _4KB) {
      ...
      md_list.hdr.list_buff_size = _4KB;
    }
    ...
    return_val = md_write_list(MD_CTX_VP, field_id, _4KB, true, true, page_list_i
== page_list_info.last_entry, md_ctx, &md_list.hdr, MD_IMPORT_MUTABLE, access_qual,
&next_field_id, tmp_ext_error_info, true);
    ...
    if (return_val != TDX_SUCCESS) {
      if (migsc_p->interrupted_state.status == TDX_SUCCESS) {
        migsc_p->interrupted_state.status = return_val;
        migsc_p->interrupted_state.extended_err_info[0] = tmp_ext_error_info[0];
        migsc_p->interrupted_state.extended_err_info[1] = tmp_ext_error_info[1];
      }
    }
    field_id = next_field_id;
    page_list_i++;
    ...
  } while ((uint64_t)page_list_i <= page_list_info.last_entry);
  ...
  if (migsc_p->interrupted_state.status != TDX_SUCCESS) {
    local_data_ptr->vmm_regs.rcx = migsc_p->interrupted_state.extended_err_info[0];
    local_data_ptr->vmm_regs.rdx = migsc_p->interrupted_state.extended_err_info[1];
    tdcs_p->management_fields.op_state = OP_STATE_FAILED_IMPORT;
    return_val = api_error_fatal(migsc_p->interrupted_state.status);
    goto EXIT;
  }
...
```

When `tdh_import_state_vp` is called by the host VMM it is passed a list of HPAs pointing to 4KB `md_list_t` structures, that make up the migration bundle, as the `page_of_list_pa` parameter (See the [Migration Bundles](#) section for additional details about the `md_list_t` structure).

After some initial validation the function proceeds to loop through each `md_list_t` structure by decrypting and copying the contents to a stack-based `md_list` buffer. The `md_list.hdr.list_buff_size` is checked to ensure it is less than or equal to `_4KB`. This is passed to `md_write_list` as `&md_list.hdr`, and with a fixed size of `_4KB`. If `md_write_list` does not return `TDX_SUCCESS` and a previous error doesn't already exist, `tmp_ext_error_info[0]` is saved to `extended_err_info[0]`. Once all `md_list` entries have been imported, `tdh_import_state_vp` checks the integrity of the migration bundle by comparing the computed MAC with the one located in the MBMD. If validation passes but an error is encountered during the import `extended_err_info` is copied to `RCX` and `RDX` and the TD `op_state` is set to `OP_STATE_FAILED_IMPORT` preventing future import activities.

```
api_error_code_e md_write_list(md_context_code_e ctx_code, md_field_id_t
expected_field, uint16_t buff_size, ..., md_context_ptrs_t md_ctx,
md_list_header_t* list_header_ptr, ...) {
    ...
    remaining_buff_size = list_header_ptr->list_buff_size - sizeof(md_list_header_t);
    sequence_buffer_ptr = (uint8_t*)(list_header_ptr) + sizeof(md_list_header_t);

    for (uint32_t i = 0; i < list_header_ptr->num_sequences; i++) {
        sequence_ptr = (md_sequence_t*)sequence_buffer_ptr;

        if (sequence_ptr->sequence_header.context_code != expected_field.context_code)
{
            ext_err_info[0] = sequence_ptr->sequence_header.raw;
            return api_error_with_l2_details(TDX_METADATA_FIELD_ID_INCORRECT, 0xFFFF,
(uint16_t)i);
        }
        ...
        retval = md_write_sequence(sequence_ptr, md_ctx, (uint32_t)remaining_buff_size,
access_type, access_qual, &elements_read, &lkp_iter, skip_non_writable,
ext_err_info, is_import);
        ...
        remaining_buff_size -= (sizeof(md_field_id_t) + (elements_read *
sizeof(uint64_t)));
        sequence_buffer_ptr += (sizeof(md_field_id_t) + (elements_read *
sizeof(uint64_t)));
        ...
```

`md_write_list` uses `list_header_ptr->num_sequences` to iterate through the sequences of an `md_list`. After performing some initial checks to make sure the `sequence_header` contains either the `expected_field` or the next non-optional field `md_write_sequence` is called.

In cases where a check fails information about the failure is returned in `ext_err_info`. For example, `sequence_header.context_code` is checked against the `expected_field.context_code` and if this check fails the raw `uint64_t` value in `sequence_ptr->sequence_header` is set to `ext_err_info[0]` and an error code constructed with `api_error_with_l2_details` is returned.

```
static api_error_code_e md_write_sequence(md_sequence_t* sequence_ptr,
md_context_ptrs_t md_ctx, uint32_t buff_size, md_access_t access_type,
md_access_qualifier_t access_qual, uint32_t* elements_read, lookup_iterator_t*
lkp_iter, bool_t skip_non_writable, uint64_t ext_err_info[2], bool_t is_import) {
```

```
    ...
    IF_RARE (buff_size < (sizeof(md_field_id_t) + sizeof(uint64_t))) {
        ext_err_info[0] = lkp_iter->field_id.raw;
        return api_error_with_l2_details(TDX_METADATA_LIST_OVERFLOW, 0xFFFF, 0);
    }
    ...
    uint32_t num_fields = sequence_ptr->sequence_header.last_field_in_sequence + 1;
    buff_size -= sizeof(md_field_id_t);

    for (uint32_t i = 0; i < num_fields; i++) {
        entry = &lkp_iter->lookup_table[lkp_iter->table_idx];
        if (sequence_ptr->sequence_header.write_mask_valid) {
            wr_mask = sequence_ptr->element[0];
            sequence_idx++;
            buff_size -= sizeof(uint64_t);
        } else { wr_mask = (uint64_t)-1; }

        if ((uint64_t)buff_size < ((uint64_t)entry->num_of_elem * sizeof(uint64_t))) {
            ext_err_info[0] = lkp_iter->field_id.raw;
            return api_error_with_l2_details(TDX_METADATA_LIST_OVERFLOW, 0xFFFF, 0);
        }

        if (!skip_non_writable || is_required_or_optional_entry(entry, access_type)) {
            retval = md_write_field_with_entry(ctx_code, lkp_iter->field_id, access_type,
access_qual, md_ctx, &sequence_ptr->element[sequence_idx], wr_mask, entry,
is_import, sequence_ptr->sequence_header.write_mask_valid);

            if (retval != TDX_SUCCESS) {
                if (!((retval == TDX_METADATA_FIELD_NOT_WRITABLE) && skip_non_writable)) {
                    ext_err_info[0] = lkp_iter->field_id.raw;
                    return retval;
                }
            }
        }
        buff_size -= (entry->num_of_elem * sizeof(uint64_t));
        sequence_idx += entry->num_of_elem;
        ...
        if ((i < (num_fields - 1)) && (is_null_field_id(lkp_iter->field_id) ||
(lkp_iter->field_id.class_code != prev_class_code))) {
            ext_err_info[0] = sequence_ptr->sequence_header.raw;
            return TDX_METADATA_FIELD_ID_INCORRECT;
        }
        ...
```

`md_write_sequence` starts by checking that `buff_size` is greater than or equal to `(sizeof(md_field_id_t) + sizeof(uint64_t)`. It then parses the `sequence_header` to capture the `num_fields` and adjusts `buff_size` before entering the loop to process the sequence elements. The loop starts by checking if the `sequence_header.write_mask_valid` flag is set and sets the `wr_mask` to `sequence_ptr->element[0]` and adjusts `buff_size`. Then there is a check to ensure `buff_size` is large enough to hold the number of elements in the entry. Finally, `md_write_field_entry` is called to write the elements in the sequence into the associated metadata field. When that returns `buff_size` is updated to reflect the remaining size by subtracting `entry->num_of_elem * sizeof(uint64_t)`.

`buff_size` passed to this function is indirectly controlled by the host VMM because `remaining_buff_size` is initialized from `list_header_ptr->list_buff_size` in `md_write_list`. The loop in `md_write_sequence` works correctly on the first iteration but subsequent iterations become problematic if `sequence_header.write_mask_valid` flag is set and `buff_size` is less than `sizeof(uint64_t)`. In this situation, `buff_size` subtracts `sizeof(uint64_t)` and causes an integer underflow. As `buff_size` is a `uint32_t` the value after the subtraction would be close to 4GB but other restrictions reduce the distance of the OOB access:

1. `sequence_header` must be a valid entry to be imported
2. `num_of_elem` is not attacker controlled and the largest had a value of 6
3. `num_fields` is computed from `last_field_in_sequence + 1` which is limited to a value between 1 and 512
4. `class_code` and `context_code` between entries must match to be in the same sequence

After some examination it was found that the largest OOB access for `tdh_import_state_vp` is achieved with `X2APIC_IDS`, `XBUFF`, `L2_MSR_BITMAPS`, `L2_MSR_BITMAPS_2`, and `L2_MSR_BITMAPS_3` fields because their `num_of_fields` value is greater than or equal to 512.

```
// XBUFF in tdvps_fields_lookup.c
.field_id = { .raw = 0x1220000300000000 },
.num_of_fields = 1536, .num_of_elem = 1, .offset = 0x3000, .attributes = { .raw =
0x0 },
.prod_rd_mask = (0ULL & 0xFFFFFFFFFFFFFFFFULL), .prod_wr_mask = (0ULL &
0xFFFFFFFFFFFFFFFFULL),
.dbg_rd_mask = (-1ULL & 0xFFFFFFFFFFFFFFFFULL), .dbg_wr_mask = (-1ULL &
0xFFFFFFFFFFFFFFFFULL),
.guest_rd_mask = (0ULL & 0xFFFFFFFFFFFFFFFFULL), .guest_wr_mask = (0ULL &
0xFFFFFFFFFFFFFFFFULL),
.export_mask = (-1ULL & 0xFFFFFFFFFFFFFFFFULL), .import_mask = (-1ULL &
0xFFFFFFFFFFFFFFFFULL),
.special_rd_handling = false, .special_wr_handling = true,
.mig_export = MIG_ME, .mig_import = MIG_ME
```

If `tdh_import_state_td` was used to perform the OOB read, the `X2APIC_IDS` field would provide the largest access.

```c
// X2APIC_IDS in tdr_tdcs_fields_lookup.c
.field_id = { .raw = 0x9C10000200000000 },
.num_of_fields = 576, .num_of_elem = 1, .offset = 0x1100, .attributes = { .raw =
0x0 },
.prod_rd_mask = (-1ULL & 0xFFFFFFFFULL), .prod_wr_mask = (0ULL & 0xFFFFFFFFULL),
.dbg_rd_mask = (-1ULL & 0xFFFFFFFFULL), .dbg_wr_mask = (0ULL & 0xFFFFFFFFULL),
.guest_rd_mask = (0ULL & 0xFFFFFFFFULL), .guest_wr_mask = (0ULL & 0xFFFFFFFFULL),
.migtd_rd_mask = (-1ULL & 0xFFFFFFFFULL), .migtd_wr_mask = (0ULL & 0xFFFFFFFFULL),
.export_mask = (-1ULL & 0xFFFFFFFFULL), .import_mask = (-1ULL & 0xFFFFFFFFULL),
.special_rd_handling = false, .special_wr_handling = false,
.mig_export = MIG_MB, .mig_import = MIG_MBO
```

There are two options for retrieving the data read from the OOB access. Regardless of the approach the greatest range is achieved by placing a `sequence_header` and `wr_mask` at the end of a sequence. Given the restrictions the range of the OOB access is limited to `(sizeof(uint64_t) * 2) * num_of_fields` or 8KB. This is achieved by having:

1. `write_mask_valid` set in the `sequence_header` each field effectively skips one element from the `wr_mask` assignment
2. `num_of_elem = 1` so each field consumes another element
3. `last_field_in_sequence` set to 511 (i.e., `num_fields` is equal to 512)

Figure 12 provides a detailed layout of how data is organized on the stack when the vulnerability is triggered.

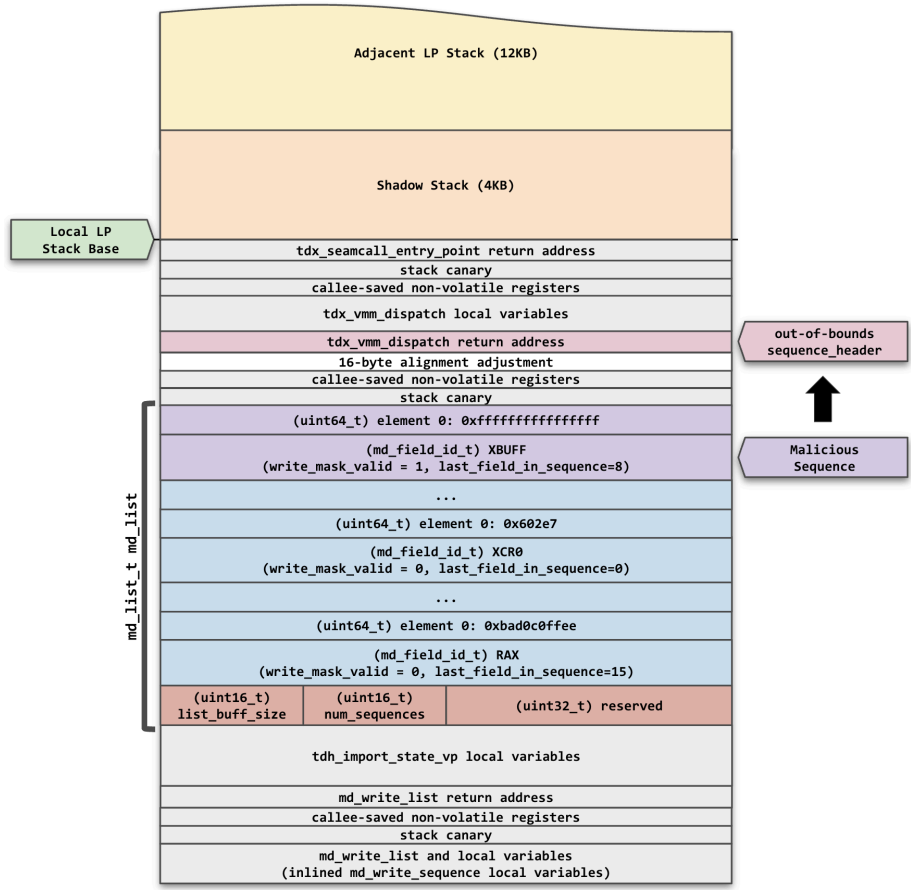

**Figure 12: Malicious Metadata List Used To Read OOB data**

**Option 1:** Use `ext_err_info[0]` to return 64bit OOB value in `RCX` to the host VMM. By aligning the end of the malicious sequence to right before the data to leak, `md_write_list` interprets the out-of-bounds `sequence_header` as the next one (See Figure 12). When the `context_code` check fails the value `sequence_ptr->sequence_header.raw` is assigned and the function returns. There is a possibility that the `context_code` would match and not fail but in this case the attacker can alternate between the `tdh_import_state_vp` and `tdh_import_state_td` API as their context codes are different (i.e., `MD_CTX_TD` is 1 and `MD_CTX_VP` is 2). The value of `i` returned in the error code and the error code itself can be used to determine if the API failed at the correct point to leak data.

```
if (sequence_ptr->sequence_header.context_code != expected_field.context_code) {
  ext_err_info[0] = sequence_ptr->sequence_header.raw;
  return api_error_with_l2_details(TDX_METADATA_FIELD_ID_INCORRECT, 0xFFFF,
(uint16_t)i);
```

**Option 2:** Because the TD being imported is under complete control of the attacker, use it to extract the OOB data from the imported state. All data from the malicious sequence to the out-of-bounds `sequence_header` will be copied into an attacker-controlled portion of `XBUFF` (See Figure 12). `num_sequences` is controlled to ensure `md_write_list` returns after completing the import of OOB data into `XBUFF`. After the import is completed `tdh_vp_enter` can be used to extract the data with a custom bootloader.

*A variation of this would be to use `X2APIC_IDS`, which is restricted to 32bit values (i.e. would only save 4 of every 8 bytes), to load the OOB data. This field could be directly read by the host VMM using the `tdh_mng_rd` API.*

ASLR is defeated because the leaked stack data includes: return address to `tdx_vmm_dispatcher`, address of `tdx_module_local_t local_data` when it is pushed to the stack by the tdh_import_state_vp prolog, and the contents of the shadow stack. Stack-smashing protection defeated because global stack canary is leaked.

LP data stacks are co-located, 12KB in size, with a 4KB shadow stack between them. The shadow stack is readable with normal memory accesses, but protected from normal write operations with Control-Flow Enforcement Technology (CET). There is no guard page to isolate an adjacent LP's stack. Even with a guard page present Option 1 wouldn't have been stopped because a `write_mask_valid` could be set with a `wr_mask` of 0 effectively skipping the accesses.

**Figure 13: Exploited Import Operation Leaking Stack Data**
*Note: An animated version of the figure is provided here*

The PoC, shown in Figure 13, performs **Option 1** using the TDXplore Toolkit. The host VMM and migTD are both compromised and cooperating to build a custom metadata bundle and correctly encrypt it for import. The leaked information is returned in `extended error information 1`, which in this case is a canonical address from the TDX module.

**Remediation:** This vulnerability was fixed by moving the block of code that checks the `valid_write_mask` and `wr_mask` to before the loop.

## Vulnerability 3: Speculative Out-of-Bounds Read in Guest RDMSR and WRMSR Handlers

**Intel Technical Advisory for CVE-2025-27572**: In certain Intel TDX modules, an out-of-bounds read condition may occur under specific microarchitectural conditions allowing speculative execution and potentially result in information exposure through side-channel analysis.

**Attack Scenario**: A malicious guest TD can execute `WRMSR` and `RDMSR` to train branch predictors and cause OOB memory accesses in the speculative execution domain, which are later processed by encoding gadgets inside the Intel TDX Module. A guest TD can use the [prime+probe](#) or [flush+flush](#) attack technique (the latter colluding with the host VMM) to extract the secret information.

The exit handlers for `WRMSR` and `RDMSR` inside the Intel TDX Module relies on the `rd_wr_msr_generic_checks` subroutine to sanity check the user input, and this subroutine has code gadgets that are vulnerable to speculative OOB reads.

The execution of certain privileged instructions inside a TD is emulated by the Intel TDX Module. In particular, when a TD executes `wrmsr`/`rdmsr` instructions, it gets interrupted by the Intel TDX Module, which emulates the execution of these instructions via `td_wrmsr_exit` and `td_rdmsr_exit`. The subroutine `rd_wr_msr_generic_checks` is responsible for checking if the requested 32-bit MSR address is within a valid range that is allowed by an internal bitmap structure:

```
// Access to any MSR not in the bitmap ranges results in a #VE
if (!is_msr_covered_by_bitmap(msr_addr)) {
  return
construct_msr_status_with_ve_category(TD_MSR_ACCESS_MSR_NON_ARCH_EXCEPTION,
VE_INFO_NON_CONFIG_PARAVIRT);
}

if ((vm_id > 0) &&
get_msr_bitmap_bit((uint8_t*)tdvps_p->l2_vm_ctrl[vm_id-1].l2_shadow_msr_bitmaps,
msr_addr, wr)) {
  return TD_MSR_ACCESS_L2_TO_L1_EXIT;
}
```

In the above code snippet, speculative execution of the conditional check for `is_msr_covered_by_bitmap` can result in invalid MSR addresses to be processed by the following subroutines. Later `get_msr_bitmap_bit` speculatively extracts a single bit of such OOB memory based on the provided MSR address, which is then used in another conditional code that depends on a single bit of leaked memory. Consequently, a malicious TD can steal a single bit of secret memory of the Intel TDX Module by probing the cache state of the second conditional code. By repeating this attack for different MSR addresses, the attacker can potentially leak the Intel TDX Module memory content one bit at a time.

**Remediation**: This vulnerability was fixed by adding an `lfence` after the OOB check in `is_msr_covered_by_bitmap`.

## Vulnerability 4: Speculative Out-of-Bounds Read in Host HKID Free and VP Flush API

**Intel Technical Advisory for CVE-2025-32467**: In some Intel TDX modules, improper initialization may lead to speculative execution behaviors, which under specific microarchitectural conditions, could result in limited information disclosure.

**Attack Scenario**: A malicious host VMM executes `tdh_mng_key_freeid` and `tdh_mng_vpflushdone` with arbitrary addresses injected into the unutilized variable on the stack. These values result in OOB access and partial information leakage, which can be recovered by flush+flush attack technique.

In various Intel TDX Module APIs, `tdr_ptr` is not initialized by default. In some `tdh_mng_key_freeid` and `tdh_mng_vpflushdone`, this uninitialized pointer can be used to access OOB data during speculative execution.
These APIs can be executed by a malicious VMM at different stages of managing a TD. The API `tdh_mng_key_freeid` is responsible for marking the guest TD's HKIDs in the Key Ownership Table (KOT) as `HKID_FREE`. Similarly, `tdh_mng_vpflushdone` is responsible for marking the TD's HKID in the KOT as `HKID_FLUSHED`.

At the beginning of these subroutines, the TDR has to be mapped using `check_lock_and_map_explicit_tdr` subroutine. A malicious VMM can provide invalid inputs to this subroutine so it fails. Upon failure, `tdr_ptr` won't be initialized, as a result picking up attacker chosen values from on the stack. Speculative execution of the check for the return code: `if (return_val != TDX_SUCCESS)` allows these subroutines to have forward progress in the speculative execution domain.

```
return_val = check_lock_and_map_explicit_tdr(tdr_pa, OPERAND_ID_RCX, TDX_RANGE_RW,
TDX_LOCK_EXCLUSIVE, PT_TDR, &tdr_pamt_block, &tdr_pamt_entry_ptr, &tdr_locked_flag,
&tdr_ptr);
if (return_val != TDX_SUCCESS) {
  TDX_ERROR("Failed to check/lock/map a TDR - error = %lld\n", return_val);
  goto EXIT;
}
```

Later, as we see in the following code snippet, an attacker who controls the `tdr_ptr` can craft arbitrary HKID values that results in accessing OOB data relative to the KOT.

```
curr_hkid = tdr_ptr->key_management_fields.hkid;

if (global_data_ptr->kot.entries[curr_hkid].wbinvd_bitmap != 0) {
  TDX_ERROR("CACHEWB is not complete for this HKID (=%x)\n", curr_hkid);
  return_val = TDX_WBCACHE_NOT_COMPLETE;
  goto EXIT;
}
```

The OOB data is later used in various conditional code blocks to leak partial information about the content of memory. This leaked data can be accessed by an attacker who is able to probe the cache state of the TDX module. For example, `kot.entries[curr_hkid].state` is checked which leak if a single byte of the stolen data is zero or not. Such partial information leak is particularly useful for attacking cryptographic keys that are mapped into the linear address space of the TDX module, which include TD private memory, which sometimes even partial information leaks can result in complete compromise of a cryptography.

Both `tdh_mng_key_freeid` and `tdh_mng_vpflushdone` expose this vulnerable code pattern when they try to update the KOT.

**Remediation:** This vulnerability was fixed by ensuring all local pointers are initialized.

## Vulnerability 5: Speculative Out-of-Bounds Read in Host API to Prebind and Bind a service TD

**Intel Technical Advisory for CVE-2025-27940**: In certain Intel TDX modules, an out-of-bounds read condition may occur under specific microarchitectural conditions may allow speculative execution and potentially result in information exposure through side-channel analysis.

**Attack Scenario**: A malicious host VMM executes `tdh_servtd_prebind` or `tdh_servtd_bind` with OOB `servtd_slot` after training the affected branch predictor, which results in OOB access followed by code gadgets that leak information about this memory access. The host VMM can use a flush+flush attack technique to recover the leaked secret.

The user input `servtd_slot` in `tdh_servtd_bind` and `tdh_servtd_prebind` APIs can bypass range checks during speculative execution and results in accessing OOB data. These APIs are responsible for binding a new service TD (currently only migration TD) to a target TD. Although currently only a single `MAX_SERVTDS` is supported, Intel TDX Module data structures support more than one service TD potential for future use cases. As a result, these APIs check if the requested `servtd_slot` is not more than the currently supported service TD (`MAX_SERVTDS` = 1).

```
if (servtd_slot >= MAX_SERVTDS) {
  return_val = api_error_with_operand_id(TDX_OPERAND_INVALID, OPERAND_ID_R8);
  goto EXIT;
}
```

However, the speculative execution of the above check allows a malicious VMM to access

OOB data based on the `servtd_bindings_table`.

```
tdcs_p->service_td_fields.servtd_bindings_table[servtd_slot]
```

This OOB memory is later processed by several branches and subroutines that may leak the value via cache. A possible scenario is this `tdx_memcmp` that would compare the OOB value against a known hash, which the attacker can construct. As a result, an attacker can learn what the OOB value is depending on the `tdx_memcmp` timing / side channel.

```
if
(!tdx_memcmp(tdcs_p->service_td_fields.servtd_bindings_table[servtd_slot].info_hash
.qwords,servtd_info_hash.qwords, sizeof(servtd_info_hash)))
```

Both `tdh_servtd_prebind` and `tdh_servtd_bind` are equally affected by this vulnerable code pattern.

**Remediation:** This vulnerability was fixed by adding a serialization instruction after the `servtd_slot` bounds checking.

# Bugs & Code Improvements

The table below lists additional bug findings, Defense in Depth (DiD) suggestions, and code improvements that have been identified throughout the review process. As Intel does not consider these items to be security vulnerabilities, they may elect to address them in one or more future sustaining releases.

| # | Type | Description |
|---|------|-------------|
| 1 | Out-of-bounds Read | Metadata list parsing leads to an integer underflow |
| 2 | Uninitialized Data Usage | Required metadata entries are skippable |
| 3 | Improper Initialization | **Illegal, stale, and unsorted TD event filter initialization** |
| 4 | Out-of-bounds Read | Out-of-bounds array indexing when locating the next entry in the CPUID lookup array |
| 5 | Missing Inline Assembly Constraints | Multiple inline assembly blocks incorrectly exclude RCX from the clobber list |
| 6 | Information Leakage | Binding handles can leak TDR HPAs for any TD |
| 7 | Denial of Service | Invalid VMCS revision identifier leads to SEAM shutdown |
| 8 | Resource Leak | HKID reservation exhaustion |
| 9 | Improper Address Check | Improper HPA and GPA checks on metadata import |
| 10 | Memory Corruption | Improper validation of host physical addresses in debug API |
| 11 | Other Spectre Gadgets | Multiple spectre gadgets allow for out-of-bounds read but can not be extracted |

## Bug 1: Metadata List Parsing Leads to an Integer Underflow

**Prerequisites**: host VMM creates a template TD to import migration bundles into. The MSK is known and the host VMM can craft a metadata bundle and correctly encrypt it.

This bug is similar to Vulnerability 2, but with a different root cause, as it can trigger an integer underflow condition by calling the `tdh_import_state_immutable`, `tdh_import_state_td`, and `tdh_import_state_vp` API with attacker crafted metadata bundles. OOB data is returned via RCX or written into TD non-memory state. `tdh_import_state_td` is used here to describe how the bug can be reached and its root cause.

```
api_error_type tdh_import_state_td(uint64_t target_tdr_pa, uint64_t
hpa_and_size_pa, uint64_t page_or_list_pa, uint64_t migs_i_and_cmd_pa) {
  ...
  md_list_header_t *md_list_hdr_p = NULL;
  ...
  md_list_t md_list;
  ...
  page_list_pa.raw = 0;
```

```
page_list_pa.page_4k_num = page_list_info.hpa;
page_list_p = (pa_t *)map_pa(page_list_pa.raw_void, TDX_RANGE_RO);
...
do {
  ...
  md_list_pa.raw = page_list_p[page_list_i].raw;
  md_list_hdr_p = (md_list_header_t *)map_pa(md_list_pa.raw_void, TDX_RANGE_RO);

  if (aes_gcm_decrypt(&migsc_p->aes_gcm_context, (uint8_t*)md_list_hdr_p,
(uint8_t*)&md_list, _4KB) != AES_GCM_NO_ERROR) {
    ...
    if (md_list.hdr.list_buff_size > _4KB) {
      ...
      md_list.hdr.list_buff_size = _4KB;
    }
    ...
  return_val = md_write_list(MD_CTX_TD, field_id, _4KB, true, true, page_list_i ==
page_list_info.last_entry, md_ctx, &md_list.hdr, MD_IMPORT_MUTABLE, access_qual,
&next_field_id, tmp_ext_error_info, true);
    ...
    if (return_val != TDX_SUCCESS) {
      if (migsc_p->interrupted_state.status == TDX_SUCCESS) {
        migsc_p->interrupted_state.status = return_val;
        migsc_p->interrupted_state.extended_err_info[0] = tmp_ext_error_info[0];
        migsc_p->interrupted_state.extended_err_info[1] = tmp_ext_error_info[1];
      }
    }
    field_id = next_field_id;
    page_list_i++;
    ...
} while ((uint64_t)page_list_i <= page_list_info.last_entry);
...
if (migsc_p->interrupted_state.status != TDX_SUCCESS) {
    local_data_ptr->vmm_regs.rcx = migsc_p->interrupted_state.extended_err_info[0];
    local_data_ptr->vmm_regs.rdx = migsc_p->interrupted_state.extended_err_info[1];
    tdcs_p->management_fields.op_state = OP_STATE_FAILED_IMPORT;
    return_val = api_error_fatal(migsc_p->interrupted_state.status);
    goto EXIT;
}
...
```

Each `md_list_t` in the metadata bundle is decrypted and copied to a stack based buffer called `md_list`. The `md_list_hdr.list_buff_size` is checked to be less than `_4KB` and then `md_write_list` is called.

```
api_error_code_e md_write_list(md_context_code_e ctx_code, md_field_id_t
expected_field, uint16_t buff_size, ..., md_context_ptrs_t md_ctx,
md_list_header_t* list_header_ptr, ...) {
   ...
   uint16_t remaining_buff_size;

   ...
   remaining_buff_size = list_header_ptr->list_buff_size - sizeof(md_list_header_t);
   sequence_buffer_ptr = (uint8_t*)(list_header_ptr) + sizeof(md_list_header_t);

   for (uint32_t i = 0; i < list_header_ptr->num_sequences; i++) {
      sequence_ptr = (md_sequence_t*)sequence_buffer_ptr;
   ...
```

`md_write_list` proceeds to initialize `remaining_buff_size`, a `uint16_t`, to `list_header_ptr->list_buff_size - sizeof(md_list_header_t)`. As `list_header_ptr->list_buff_size` was never checked to be greater than or equal to `sizeof(md_list_header_t)` an integer underflow could occur. `md_list_header_t` has a size of 8 bytes so a value less than this, such as 0, causes `remaining_buff_size` to be initialized to a value close to 64KB, which is greater than the 4KB stack allocated `md_list` pointed to by `list_header_ptr`.

Each sequence is parsed by `md_write_sequence` and can contain a maximum of 512 fields making the maximum size `(512 * sizeof(uint64_t)) + sizeof(sequence_header)` or 4104 bytes.

In addition to returning OOB data via `RCX` to the host VMM or writing it into the TD non-memory state, 8 bytes of stale stack data could be accessible. This is because during the prolog of `tdh_import_state_immutable`, `tdh_import_state_td`, and `tdh_import_state_vp` an alignment operation is performed before allocating space for local variables (i.e. `AND RSP, -0x10`, as shown in Figure 14).

**Figure 14: Stale or Uninitialized Stack Locations**

`tdx_vmm_dispatcher` performs the same alignment operation, allocates `0x20` bytes of storage for local variables and then calls other API, such as `tdh_import_state_immutable`. The stack upon entry to the callee is only 8 byte aligned because of the return address pushed to the stack by the `call` instruction. The callee prolog then pushes six registers (i.e. `RBP`, `R15`, `R14`, `R13`, `R12`, and `RBX`) to the stack leaving `RSP` only 8 byte aligned. The align operation is then performed effectively allocating 8 bytes of uninitialized data on the stack.

Before leaking out-of-bounds data with this bug the LP could have interacted with another TD populating the 8 byte uninitialized region with what would become accessible stale data.

Implications of this stale stack data were not investigated further to understand if TD specific confidential information would be present. Even without this the bug breaks ASLR, leaks the global stack canary, and the contents of a shadow stack in the Intel TDX Module.

This bug was not classified as a security vulnerability by Intel's Product Security & Incident Response Team (PSIRT) and not assigned a CVE identifier. The leakable data is confined to the current LP data stack and adjacent LP shadow stack. Additionally, Intel considers ASLR and stack canaries, especially when CET is enabled, DiD mechanisms.

**Remediation**: This bug was fixed in the updated Intel TDX Module.

## Bug 2: Required Metadata Entries are Skippable

**Prerequisites**: A host VMM creates a template TD to import migration bundles into. The host VMM is given the MSK, crafts a metadata bundle, and encrypts it for import.

`tdh_import_state_immutable`, `tdh_import_state_td`, and `tdh_import_state_vp` call `md_write_list` with `skip_non_writable` set to `true`. For each sequence in the metadata list `md_write_sequence` is called. A sequence can specify, via the `write_mask_valid` flag in the `sequence_header`, whether `element[0]` holds a `wr_mask`.

```c
static api_error_code_e md_write_sequence(md_sequence_t* sequence_ptr, ..., bool_t
skip_non_writable,
  ...
    if (sequence_ptr->sequence_header.write_mask_valid) {
      wr_mask = sequence_ptr->element[0];
      ...
    }
    ...
    if (!skip_non_writable || is_required_or_optional_entry(entry, access_type)) {
      retval = md_write_field_with_entry(ctx_code, lkp_iter->field_id, access_type,
access_qual, md_ctx, &sequence_ptr->element[sequence_idx], wr_mask, entry,
is_import, sequence_ptr->sequence_header.write_mask_valid);
      if (retval != TDX_SUCCESS) {
        if (!((retval == TDX_METADATA_FIELD_NOT_WRITABLE) && skip_non_writable)) {
          ext_err_info[0] = lkp_iter->field_id.raw;
```

```
        return retval;
    }
...
```

If `write_mask_valid` is `true` and `element[0]` is `0` then `md_write_field_with_entry` returns early with `TDX_METADATA_FIELD_NOT_WRITABLE`. When `skip_non_writable` is `true` the field is skipped and no error is returned allowing *any* required entry to be skipped.

```
api_error_code_e md_vp_write_element(md_field_id_t field_id, const md_lookup_t*
entry, md_access_t access_type,
  ...
  if (combined_wr_mask == 0) {
      return TDX_METADATA_FIELD_NOT_WRITABLE;
  }
  ...
```

Processing required entries is extremely important because some have special handling, perform verification, do initialization, and in some cases are assumed valid when subsequent entries are imported. When these entries are skipped, they are left in their `tdh_mng_add_cx` initialized state. Because verification and initialization are performed as entries are imported there are multiple instances where entries are not checked when the overall import completes. For example, the `MD_TDCS_EPTP` is initialized in `md_td_write_field` because it has `special_wr_handling` set to `true`.

```
// EPTP // 19
.field_id = { .raw = 0x1110000300000004 },
.num_of_fields = 1, .num_of_elem = 1, .offset = 0x0098, .attributes = { .raw = 0x0
},
.prod_rd_mask = (-1ULL & 0xFFFFFFFFFFFFFFFFULL), .prod_wr_mask = (0ULL &
0xFFFFFFFFFFFFFFFFULL),
.dbg_rd_mask = (-1ULL & 0xFFFFFFFFFFFFFFFFULL), .dbg_wr_mask = (0ULL &
0xFFFFFFFFFFFFFFFFULL),
.guest_rd_mask = (0ULL & 0xFFFFFFFFFFFFFFFFULL), .guest_wr_mask = (0ULL &
0xFFFFFFFFFFFFFFFFULL),
.migtd_rd_mask = (18442240474082185215ULL & 0xFFFFFFFFFFFFFFFFULL), .migtd_wr_mask
= (0ULL & 0xFFFFFFFFFFFFFFFFULL),
.export_mask = (18442240474082185215ULL & 0xFFFFFFFFFFFFFFFFULL), .import_mask =
(18442240474082185215ULL & 0xFFFFFFFFFFFFFFFFULL),
.special_rd_handling = false, .special_wr_handling = true,
.mig_export = MIG_MB, .mig_import = MIG_MB
```

When imported this calls `verify_and_set_td_eptp_controls` to validate and set the
`tdcs_ptr->executions_ctl_fields.eptp` field.

```
api_error_code_e md_td_write_field(md_field_id_t field_id, const md_lookup_t*
entry,md_access_t access_type,
  ...
  if (combined_wr_mask == 0) {
   return TDX_METADATA_FIELD_NOT_WRITABLE;
  }
  ...
    case MD_TDCS_EPTP_FIELD_ID:
      ...
      eptp.raw = value[0] & combined_wr_mask;
      if (!verify_and_set_td_eptp_controls(md_ctx.tdr_ptr, md_ctx.tdcs_ptr,
md_ctx.tdcs_ptr->executions_ctl_fields.gpaw, eptp)) {
        return TDX_METADATA_FIELD_VALUE_NOT_VALID;
      }
      write_done = true;
      break;
      ...
```

As a side note, this function is also responsible for validating that
`tdcs_ptr->executions_ctl_fields.gpaw` is consistent with the state of `ept_pwl`.
`tdcs_ptr->executions_ctl_fields.gpaw` and `tdcs_ptr->executions_ctl_fields.eptp`
could be configured inconsistently by first setting the `gpaw` to `true` and the `ept_pwl` to
`LVL_PML5` and then restarting the immutable import and setting `gpaw` to `false` and skipping
the `MD_TDCS_EPTP` field.

```
bool_t verify_and_set_td_eptp_controls(tdr_t* tdr_ptr, tdcs_t* tdcs_ptr, bool_t
gpaw, ia32e_eptp_t eptp)
  ...
  if (gpaw && (eptp.fields.ept_pwl < LVL_PML5)) {
    return false;
  }

  tdcs_ptr->executions_ctl_fields.gpaw = gpaw;
  ...
  pa_t sept_root_pa;
  sept_root_pa.raw = tdr_ptr->management_fields.tdcx_pa[SEPT_ROOT_PAGE_INDEX];

  eptp.fields.base_pa = sept_root_pa.page_4k_num;

  tdcs_ptr->executions_ctl_fields.eptp.raw = eptp.raw;
```

```
  return true
}
```

When skipped, `tdcs_ptr->execution_ctl_fields.eptp` is left with its initialization value of `SEPTE_L2_INIT_VALUE`, which is `0`, from `tdh_mng_add_cx`. Once `tdh_import_state_immutable` completes, the `op_state` is switched to `OP_STATE_MEMORY_IMPORT` and API to perform SEPT walks are allowed.

A `tdcs_ptr->execution_ctl_fields.eptp` of `0` is interpreted as:

1. `ept_ps_mt` - paging-structure memory type of `MT_UC`
2. **`ept_pwl` - page-walk length of `LVL_PT`**
    a. VMEntry requires the value to be `LVL_PML5` or `LVL_PML4`
3. `enable_ad_bits` - set accessed and dirty flags is `false`
4. `enable_sss_control` - supervisor shadow stack control is `false`
5. **`base_pa` - physical address of the root paging structure as 0**

```
ia32e_sept_t* secure_ept_walk(ia32e_eptp_t septp, pa_t gpa, uint16_t private_hkid,
...
  ia32e_sept_t *pte;
...
  ept_level_t requested_level = *level;
  ept_level_t current_lvl;
...
  pt_pa.raw = septp.raw & IA32E_PAGING_STRUCT_ADDR_MASK;
  current_lvl = septp.fields.ept_pwl;
  for (;current_lvl >= LVL_PT; current_lvl--) {
    pt_pa = set_hkid_to_pa(pt_pa, private_hkid);
    pt = map_pa((void*)(pt_pa.full_pa), TDX_RANGE_RW);
    pte = &(pt->sept[get_ept_entry_idx(gpa, current_lvl)]);

    cached_sept_entry->raw = pte->raw;
    *level = current_lvl;

    if (current_lvl == requested_level) {
      break;
    }
    ...
  }
  return pte;
}
```

`secure_ept_walk` maps HPA 0 with the TD HKID and proceeds to try and walk the SEPT table.

`current_lvl` is initialized from `septp.fields.ept_pwl` to `LVL_PT` when `pte` is dereferenced a Machine Check Exception (#MCE) is triggered because private memory wasn't properly initialized, leading to a SEAM shutdown.

The table below provides additional fields that are non-optional (i.e. have a `mig_import` value of `MIG_MB` or `MIG_ME`), are skippable because of this bug, and would leave a TDCS or TDVPS fields in invalid states. While other fields beyond this table can be skipped, their pre-initialized value is either valid or the field has no effect (e.g., a `mig_import` value of `MIG_CB`). Beyond the `MD_TDCS_EPTP` field discussed above only the `MD_TDVPS_XCR0` was found to have impact beyond that of the TD being imported.

| Field Identifier | Description |
|---|---|
| `MD_TDVPS_XCR0` | `guest_state.xcr0` when skipped has a value of 0 but expects `x87_fpu_mmx` (i.e., bit 0) to be set. This is checked by `check_guest_xcr0_value` called when the field is written.<br><br>`tdvps_ptr->guest_state.xcr0` is loaded by `ia32_xsetbv` in `restore_guest_td_state_before_td_entry`. This is called from `tdh_vp_enter` and would raise a #GP(0) exception leading to a SEAM shutdown. |
| `MD_TDCS_NUM_VCPUS` | `management_fields.num_vcpus` when skipped has a value of 0. When written during an import by `md_td_write_field` this value is verified to be greater than 0 and less than `MAX_VCPUS_PER_TD`.<br><br>An imported TD could be switched to the `OP_STATE_POST_IMPORT` state without any imported VPs. `tdh_import_track` only checks that `tdcs_p->migration_fields.num_migrated_vcpus != tdcs_p->management_fields.num_vcpus`. |
| `MD_TDCS_TSC_FREQUENCY` | `executions_ctl_fields.tsc_frequency` when skipped has a value of 0. When written during an import by `md_td_write_field` this value is verified to be between `VIRT_TSC_FREQUENCY_MIN` (4) and `VIRT_TSC_FREQUENCY_MAX` (400). This value is later used by `calculate_tsc_virt_params` when the `MD_TDCS_VIRTUAL_TSC` field is imported. |
| `MD_TDCS_VIRTUAL_TSC` | This field when skipped leaves `tsc_multiplier` and `tsc_offset` with values of 0. This is later used by `calculate_virt_tsc` when `MD_TDVPS_TSC_DEADLINE` fields are imported. |
| `MD_TDCS_HP_LOCK_TIMEOUT` | `executions_ctl_fields.hp_lock_timeout` when skipped has a value of 0. When written during an import by `md_td_write_field` this value is verified to be greater than between `MIN_HP_LOCK_TIMEOUT_USEC` (10000) and `MAX_HP_LOCK_TIMEOUT_USEC` (100000000). |
| `MD_TDCS_EXPORT_COUNT` | `migration_fields.export_count` has a `.import_mask` of `(-1ULL & 0xFFFFFFFFULL)` meaning any 32-bit value is allowed. `MAX_EXPORT_COUNT` as checked in `tdh_export_state_immutable` expects the value to be less than `MAX_EXPORT_COUNT` (0x7FFFFFFF) but import does not provide a similar check or constraint. |

The `MD_TDCS_EPTP` and `MD_TDVPS_XCR0` were found to induce SEAM shutdowns and variants from the table above were not exhaustively analyzed.

Exploitation requires a compromised host VMM and migTD (both non-Intel managed components) working in cooperation to construct, encrypt, and transfer metadata bundles to the Intel TDX Module.

As the threat model for Intel TDX does not include Availability, this bug was not classified as a security vulnerability by Intel PSIRT.

**Remediation**: Intel has confirmed the bug, and indicated it would be fixed in a future release.

## Bug 3: Illegal, Stale, and Unsorted TD Event Filter Initialization

**Prerequisites**: The Intel TDX Module has been configured and initialized on the platform.

The `tdh_mng_init` API is used during the [Build Sequence](#) when creating a TD. It initializes the global-scope of the TD and TD-scope state shared by its VPs. It's callable with a TD `op_state` of `OP_STATE_UNINITIALIZED` and only changes the state to `OP_STATE_INITIALIZED` on success. In the case where an error is called the TD is left in `OP_STATE_UNINITIALIZED`.

```
api_error_type tdh_mng_init(uint64_t target_tdr_pa, uint64_t target_td_params_pa,
uint64_t event_filters_info_params) {
    ...
    bool_t event_filtering = target_tdr_pa & BIT(0);
    ...
    event_filter_info_t event_filters_info = { .raw = event_filters_info_params };

    if (event_filtering && tdcs_ptr->executions_ctl_fields.attributes.perfmon) {
        ...
        tdcs_ptr->executions_ctl2_fields.event_filters_num =
event_filters_info.event_filters_num;

        event_filters_p = (event_filter_t*)map_pa((void*)(event_filters_info.raw &
~BITS(11, 0)), TDX_RANGE_RO);

        for (uint16_t i = 0; i < event_filters_info.event_filters_num; i++) {
            event_filter_t event_filter = event_filters_p[i];
            if (event_filter.reserved_0 || event_filter.umask > 0xFF ||
event_filter.negative || event_filter.umask_mask != 0xFFFF) {
                TDX_ERROR("Illegal event filter [%d] = 0x%lx\n", i, event_filter.raw);
                return_val = api_error_with_operand_id(TDX_EVENT_FILTER_INVALID, i);
                goto EXIT;
            }

            event_filter_internal.event_select = (uint8_t)event_filter.event_select;
            event_filter_internal.umask = (uint8_t)event_filter.umask;

            if ((i != 0) && (tdcs_ptr->event_filters_internal[i - 1].raw >=
event_filter_internal.raw)) {
                TDX_ERROR("Event filters array must be sorted\n");
                return_val = api_error_with_operand_id(TDX_EVENT_FILTER_ORDER_INVALID, i);
                goto EXIT;
            }

            tdcs_ptr->event_filters_internal[i] = event_filter_internal;
        }
        ...
```

This API accepts a parameter called `event_filters_info_params` which is used to initialize the `tdcs_ptr->event_filters_internal` array when `event_filtering` is `true` and the `perfmon` flag is set.

The initialization loop checks to ensure that each `event_filter` is supported and the array is sorted. Neither condition is correctly enforced and `tdcs_ptr->executions_ctl2_fields.event_filters_num` can be incorrectly initialized.

The `event_filters_internal` array and `event_filters_num` are later used by `is_event_allowed` and performs a binary search of the array.

Multiple calls to `tdh_mng_init` can be used to allow illegal, stale, and unsorted data into the `event_filters_internal` array. Because `event_filters_num` is assigned before processing entries, an illegal event filter would cause a check in the loop to fail resulting in a `goto EXIT;` statement being executed. `event_filters_num` is not reset and already processed event filters are left with their initialized state.

To do this, `n` calls could be crafted to fail after partially initializing fewer event filter entries each time. The next call could set `event_filters_num` to a value greater than what was initialized by any prior call, but less than `MAX_EVENT_FILTERS`, and fail. The last call could completely skip event filter initialization by setting `event_filtering` to `false`. In this situation:

1. Stale event filters exist because entries from prior calls are not removed
2. Unsorted event filters exist because `event_filters_num` was set to include the stale entries which didn't didn't pass the sort check
3. Illegal event filters exist because `event_filters_num` was set to include uninitialized entries and would hold an initialization value of `SEPTE_L2_INIT_VALUE`, which is 0, set by `tdh_mng_add_cx` call

Even though the `event_filters_internal` array is used later, these conditions did not lead to an exploitable condition for a few reasons:

1. `event_filters_num` was constrained to be less than `MAX_EVENT_FILTERS`
2. `wrmsr_ia32_perfevtsel` which calls `is_event_allowed` checks to make sure `event_filters_num` is not equal to zero
3. `is_event_allowed` can't go OOB and only uses the entries for a comparison

**Remediation:** Intel has confirmed the bug, categorized it as a functional issue, and is fixing it in a future release.

## Bug 4: Out-of-Bounds Array Indexing When Locating the Next Entry in the CPUID Lookup Array

**Prerequisites**: A TD is in an `op_state` where the metadata APIs are accessible from the host VMM.

The `md_get_next_cpuid_value_entry` function is used to find the next valid entry in the `cpu_lookup` array. This function can be called by by `md_get_next_item_with_iterator` when the `context_code` is `MD_CTX_TD` and `class_code` of `MD_TDCS_CPUID_CLASS_CODE` for the `lookup_context->field_id`. At a higher level this function is used by `md_write_list` and `md_dump_list` during the import and export of TD non-memory state.

```
const cpuid_lookup_t cpuid_lookup[MAX_NUM_CPUID_LOOKUP] = {
...
  [78] = { .leaf_subleaf = {.leaf = 0x80000002, .subleaf = 0xffffffff},
  .valid_entry = true,
  .fixed1 = { .eax = 0x65746e49, .ebx = 0x58204454, .ecx = 0x6c202020 },
  .fixed0_or_dynamic = { .eax = 0x9a8b91b6, .ebx = 0xa7dfbbab, .ecx = 0x93dfdfdf,
.edx = 0xffffffff },
  .config_index = CPUID_CONFIG_NULL_IDX
  },
...
};
```

The `cpuid_lookup` array has a size of `MAX_NUM_CPUID_LOOKUP` or 79 entries. The last entry in the array is present with the `valid_entry` field set to `true`.

```
static md_field_id_t md_get_next_cpuid_value_entry(md_field_id_t field_id, bool_t
element) {
  ...
  uint32_t leaf, subleaf;
  md_cpuid_field_id_get_leaf_subleaf(field_id, &leaf, &subleaf);
  uint32_t index = get_cpuid_lookup_entry(leaf, subleaf);

  do {
    index = index + 1;
  } while (!cpuid_lookup[index].valid_entry);

  IF_RARE (index >= MAX_NUM_CPUID_LOOKUP) {
    return (md_field_id_t)MD_FIELD_ID_NA;
  }
  ...
```

`md_get_next_cpuid_entry` uses `get_cpuid_lookup_entry` to get the `index` to start looking for the next valid entry from. If `leaf` was `0x80000002` and `subleaf` was `0xffffffff` then `78` would be returned and used to initialize `index`. The loop starts by incrementing `index` and then checking to see if `valid_entry` is `true` and continues until it is reached. Once the next entry is found, `index` is checked to make sure it's less than `MAX_NUM_CPUID_LOOKUP` but at this point the loop already performed one or more out-of-bounds indexing and dereference operations to get the value of `valid_entry`.

**Remediation:** Analysis concluded the bug is unexploitable and there is no impact to the overall security of the Intel TDX Module. Intel was aware of the issue when reported and is fixing it in an upcoming release.

## Bug 5: Multiple Inline Assembly Blocks Incorrectly Exclude RCX from the Clobber List

**Prerequisites**: Not Applicable.

The `REP` and `REPE` prefix are used to repeat an instruction which implicitly uses `RCX`, `ECX`, or `CX` as a counter to indicate how many times to repeat an instruction. This counter is decremented with each iteration of the instruction. Multiple API on the Intel TDX module and SEAMLDR improperly exclude this implicit register from the clobber list, which under certain circumstances and compiler optimizations could lead to memory corruption.

```
_STATIC_INLINE_ void tdx_memcpy(void * dst, uint64_t dst_bytes, void * src,
uint64_t nbytes) {
  volatile uint64_t junk_a, junk_b;

  tdx_sanity_check (dst_bytes >= nbytes, FATAL_ERROR_ID_183, 1);

  _ASM_VOLATILE_ ("rep; movsb;"
                  :"=S"(junk_a), "=D"(junk_b)
                  :"c"(nbytes), "S"(src), "D"(dst)
                  :"memory");
}
```

From the Intel TDX module `tdx_memcpy`, `tdx_memcmp`, and `basic_memset` are affected. From the SEAMLDR `pseamldr_memcpy`, `pseamldr_memcmp`, `basic_memset` are affected. Both are built using `clang` with the Intel TDX module using `-Os` and the SEAMLDR using `-O2` for optimization. Both optimizations can exhibit the incorrect re-use after modification by inline assembly. `-O1` also exhibits the issue while `-O0` or not specifying an optimization does not. For `-O0` this is because the `tdx_memcpy` function isn't inlined and is instead executed via the `CALL` instruction.

These functions do not include either `%rcx` in the clobber list or `=c({variable})` in the output list to indicate that `RCX` will be modified during the inline assembly sequence.

The exact compiled code sequence leading to an exploitable situation was not observed in the Intel TDX module or SEAMLDR. If unfixed future changes to the affected API, their usage, or changes in how the compiler performs optimization could manifest as memory corruption.

Variant analysis was performed with Gemini and discussed further in the [Code Difference](#) section. [Appendix B](#) is a source code PoC showing how the issue could have manifested.

**Remediation:** Intel was aware of the issue when reported. The Intel TDX module fix has been made available, and the SEAMLDR aspect will be addressed in a future release.

## Bug 6: Binding Handles Can Leak TDR HPAs for any TD

**Prerequisites**: A guest TD can interact with the Intel TDX module.

A TD can leak the TDR HPA for *any* TD by calling `tdg_servtd_rd` and `tdg_servtd_wr` API and checking potential HPAs. Different error codes are returned depending on if the provided HPA is associated with a TDR or not. While these API are meant for service TDs the checks to know if the service TD is bound to a target TD can't occur until after the TDR is located.

```
typedef union servtd_binding_handle_u {
  struct {
    uint64_t binding_slot : 12;
    uint64_t tdr_page     : 40;
    uint64_t reserved     : 12;
  };
  uint64_t raw;
} servtd_binding_handle_t;
```

The `servtd_binding_handle_t` holds a `binding_slot` and the `tdr_hpa` for the target TD. This value is used by the service TD when communicating with the Intel TDX module through the `tdg_servtd_rd` and `tdg_servtd_wr` APIs. Normally, the `servtd_binding_handle_t` is returned by `tdh_servtd_bind`, to the host VMM, and provided to a service TD so it can interact with metadata of a target TD.

```
_STATIC_INLINE_ void break_servtd_binding_handle(servtd_binding_handle_t handle,
uint256_t servtd_uuid, pa_t* tdr_hpa, uint64_t* slot) {
  handle.raw -= servtd_uuid.qwords[0];
  tdr_hpa->raw = 0;
  tdr_hpa->page_4k_num = handle.tdr_page;
  *slot = handle.binding_slot;
}
```

`break_servtd_binding_handle` is used to extract the `tdr_hpa` and `slot` values from the handle. The `servtd_uuid` is the Universally Unique Identifier (UUID) of the service TD and can be retrieved with a call to `tdg_vm_rd`.

```
static api_error_type tdg_servtd_rd_wr(servtd_binding_handle_t binding_handle,
md_field_id_t field_id, bool_t write, uint64_t wr_value, uint64_t wr_request_mask)
{
  ...
  break_servtd_binding_handle(binding_handle,
lp->vp_ctx.tdr->management_fields.td_uuid, &target_tdr_pa, &target_slot);
  ...
  return_val = othertd_check_lock_and_map_explicit_tdr(target_tdr_pa,
OPERAND_ID_TDR, write ? TDX_RANGE_RW : TDX_RANGE_RO, TDX_LOCK_SHARED, PT_TDR,
&target_tdr_pamt_block, &target_tdr_pamt_entry_ptr,
&target_tdr_locked_flag,&target_tdr_ptr);

  if (return_val != TDX_SUCCESS) {
    if (is_operand_busy_error_code(return_val)) {
      TDX_ERROR("Failed to check/lock/map a Target TDR - error = %llx\n",
return_val);
      goto EXIT;
    }
    ...
    else {
      cross_td_trap_status = return_val;
      goto EXIT;
    }
  }
...
  if
(!is_equal_256bit(target_tdcs_ptr->service_td_fields.servtd_bindings_table[target_s
lot].uuid, lp->vp_ctx.tdr->management_fields.td_uuid)) {
    cross_td_trap_status = TDX_SERVTD_UUID_MISMATCH;
    goto EXIT;
  }
...
```

This explicit usage of HPAs by a TD to interact with the Intel TDX Module provides unnecessary visibility into the layout of the host physical address space that wouldn't otherwise be accessible.

**Remediation**: Restricting exposure of host physical addresses to a TD is not a security requirement of Intel TDX. Intel is tracking this as an architectural issue and is considering the return of a generic error code for this failure case.

## Bug 7: Invalid VMCS revision identifier leads to SEAM shutdown

**Prerequisites**: The Intel TDX Module has been configured and initialized on the platform.

A host VMM can use the `VMPTRLD` instruction to load an HPA of a Virtual Machine Control Structure (VMCS) into the CPU. When this instruction executes it first validates that the HPA is 4KB aligned, bits beyond the processor's physical address width are 0, and the revision identifier is set accordingly. When these checks pass, the HPA is loaded into a special location known as the `current-VMCS` inside the LP. If a failure occurs, `current-VMCS` is not loaded and the VM failure condition is reflected in `RFLAGS`.

The revision identifier is stored in bits 30:0 at byte offset 0 in the VMCS; bit 31 is the shadow-VMCS indicator. Software is required to set this to the value of bits 30:0 in the `IA32_VMX_BASIC` MSR.

When `SEAMCALL` is executed by a host VMM the transition into SEAM is similar to a VM Exit. SEAM uses a Transfer VMCS to save the state of the host VMM into the guest fields and loads the state of SEAM from the host fields. As the host VMM is able to manage legacy VMs in addition to using the Intel TDX Module to run TDs the `current-VMCS` field is saved into the `VMCS link pointer` field in the SEAM Transfer VMCS.

When `SEAMRET` is executed, the host VMM state is restored from the SEAM Transfer VMCS guest state, similar to a VM Entry. Restoration involves loading the `current-VMCS` with the HPA saved into the `VMCS link pointer` field. However, the HPA is first checked, much like when `VMPTRLD` is executed. If this check fails, it triggers a failed VM Entry VM Exit and causes the Intel TDX Module to enter a fatal error state, which leads to a SEAM shutdown.

To trigger a SEAM shutdown, the host VMM can load a correctly initialized VMCS using `VMPTRLD`, modify the revision identifier to be invalid, and execute `SEAMCALL`.

The TDX threat model does not include Availability, so this issue doesn't impact security and wasn't assigned a CVE by Intel's PSIRT.

**Remediation**: Intel does not plan on addressing this behavior because a malicious host could just as easily shut the system down or block the usage of TDX.

## Bug 8: HKID Reservation Exhaustion

**Prerequisites**: The Intel TDX Module is loaded but has not yet been configured.

Before the Intel TDX Module can be used to run TDs the host VMM must initialize it using a sequence of API calls described in section 3.1.1 of the Intel Trust Domain Extensions (Intel TDX) Module Base Architecture Specification. `tdh_sys_config` is required to be called once from a single LP and is used to configure the Trust Domain Memory Range (TDMR), Physical Attribute Metadata Table (PAMT), and reserve the HKID to be used by the Intel TDX Module.

```
api_error_type tdh_sys_config(uint64_t tdmr_info_array_pa, uint64_t
num_of_tdmr_entries, sys_config_options_t sysconfig_options) {
  ...
  tdx_global_data_ptr->kot.entries[hkid].state = KOT_STATE_HKID_RESERVED;
  tdx_global_data_ptr->hkid = hkid;

  tdmr_pa_array = map_pa(tdmr_info_pa.raw_void, TDX_RANGE_RO);
  ...
  for(uint64_t i = 0; i < num_of_tdmr_entries; i++) {
    tdmr_entry.raw = tdmr_pa_array[i];
    retval = shared_hpa_check_with_pwr_2_alignment(tdmr_entry,
TDMR_INFO_ENTRY_PTR_ARRAY_ALIGNMENT);
    if (retval != TDX_SUCCESS) {
      retval = api_error_with_operand_id(retval, OPERAND_ID_RCX);
      TDX_ERROR("TDMR entry PA is not a valid shared HPA pa=0x%llx,
error=0x%llx\n", tdmr_entry.raw, retval);
      goto EXIT;
    }
    ...
EXIT:

  if (global_lock_acquired) {
    release_sharex_lock_ex(&tdx_global_data_ptr->global_lock);
  }

  if (tdmr_info_p_init) {
    free_la(tdmr_pa_array);
  }
  return retval;
}
```

The specific HKID to use is specified by the host VMM through the `sysconfig_options` parameter. After some validation checking the KOT entry for the HKID is set to `KOT_STATE_HKID_RESERVED`. Next the TDMR entries are processed and if an error occurs the `goto EXIT;` statement is executed which releases and frees resources. The HKID entry in the KOT is never restored to the `KOT_STATE_HKID_FREE` state. Multiple calls to this API could be used to set all entries in the KOT to `KOT_STATE_HKID_RESERVED` and the Intel TDX Module after completing the initialization sequence would be unable to run a TD.

This API is only available during the Intel TDX Module initialization sequence and the issue is isolated to being able to exhaust the KOT.

**Remediation:** Intel has confirmed the bug, categorized it as a functional issue, and is fixing it in a future release.

## Bug 9: Improper HPA and GPA Checks on Metadata Import

**Prerequisites**: Not Applicable.

Multiple metadata fields imported during a TD migration represent Guest Physical Addresses (GPAs) for the TD. `md_vp_handle_field_attribute_on_wr` is used to ensure that both HPAs and GPAs are correct based on attributes associated with the metadata entry.

```c
static api_error_code_e md_vp_handle_field_attribute_on_wr(md_field_id_t field_id,
const md_lookup_t* entry, md_context_ptrs_t md_ctx, md_access_t access_type,
uint64_t* wr_value) {
  if (entry->attributes.hpa && entry->attributes.shared) {
    uint64_t size = md_vp_get_checked_size_of_shared_hpa_range(field_id);

    if (MD_IMPORT_IMMUTABLE != access_type && MD_IMPORT_MUTABLE != access_type &&
*wr_value != NULL_PA && shared_hpa_check((pa_t)*wr_value, size) != TDX_SUCCESS) {
      return TDX_METADATA_FIELD_VALUE_NOT_VALID;
    }
  }
  else if (entry->attributes.gpa && entry->attributes.prvate) {
    if (MD_IMPORT_IMMUTABLE != access_type && MD_IMPORT_MUTABLE != access_type &&
*wr_value != NULL_PA && !check_gpa_validity((pa_t)*wr_value,
md_ctx.tdcs_ptr->executions_ctl_fields.gpaw, PRIVATE_ONLY,
md_ctx.tdcs_ptr->executions_ctl_fields.virt_maxpa)) {
      return TDX_METADATA_FIELD_VALUE_NOT_VALID;
    }
  }
  return TDX_SUCCESS;
}
```

This works as expected unless the `access_type` is `MD_IMPORT_IMMUTABLE` or `MD_IMPORT_MUTABLE`, which is the case during import operations. In this case both `check_gpa_validity` and `shared_hpa_check` are skipped and `TDX_SUCCESS` is returned.

There are a few importable fields that would match `.attributes = { .raw = 0x6}`, which is `gpa` and `prvate`. Specifically, Virtual-APIC address, HLAT pointer, and PDPTEn from the `td_l2_vmcs_fields_lookup.c` would allow for writing a GPA without the `check_gpa_validity` check. Later checks in `tdh_vp_enter` would raise EPT violations. There are no importable fields that would match `.attributes = { .raw = 0x9}` which is `hpa` and `shared` because the `.import_masks` are effectively 0. This makes sense as HPAs as different platforms would instead allocate resources and perform initialization.

**Remediation:** Intel has confirmed the bug, categorized it as a functional issue, and is fixing it in a future release.

## Bug 10: Improper Validation of Host Physical Addresses in Debug API

**Prerequisites**: Platform is running a debug version of the Intel TDX Module, which is **_only_** possible on Intel TDX development systems.

A type confusion bug exists in the Intel TDX Module's debug printing mechanism when `DEBUGFEATURE_TDX_DBG_TRACE` is enabled and was identified independently through manual analysis and with Gemini.

```
uint64_t td_debug_config(uint64_t leaf, uint64_t payload, uint64_t second_payload)
{
  ...
  if (leaf == 0) { // Set debug print target
    print_target_e print_target = (print_target_e)payload;
    debug_message_t* target_buffer = NULL;
    if (print_target == TARGET_EXTERNAL_BUFFER) {
      if (second_payload % MAX_PRINT_LENGTH) { // Check alignment
        return TDX_OPERAND_INVALID;
      }
      target_buffer = (debug_message_t*)second_payload;
    }
    ...
    p_ctl->print_target = print_target;
    p_ctl->trace_buffer = target_buffer;
    ...
}
```

A host VMM can configure the print target to `TARGET_EXTERNAL_BUFFER` via `td_debug_config`, by specifying the HPA that will be used for the `p_ctl->trace_buffer` field in the global `debug_control_t` structure.

```
uint32_t dump_print_buffer_to_vmm_memory(uint64_t hpa, uint32_t
num_of_messages_from_the_end) {
  ...
  while (reader_pos != p_ctl->buffer_writer_pos) {
    char* msg_buf_ptr = p_ctl->trace_buffer[reader_pos].message;
    vmm_buf_pos += dump_message_to_vmm_memory(msg_buf_ptr, hpa + vmm_buf_pos,
MAX_PRINT_LENGTH);
    reader_pos = get_advanced_reader_pos(reader_pos, 1);
  }
  ...
```

```
}

_STATIC_INLINE_ uint32_t dump_message_to_vmm_memory(char* msg_buf_ptr, uint64_t
hpa, uint32_t len) {
  uint32_t vmm_buf_pos = 0;
  char* vmm_buf_ptr = map_pa((void*)hpa, TDX_RANGE_RW);

  for (uint32_t i = 0; msg_buf_ptr[i] != 0 && i < len; i++) {
    if ((hpa + vmm_buf_pos) % PAGE_SIZE_IN_BYTES == 0) {
      free_la(vmm_buf_ptr);
      vmm_buf_ptr = map_pa((void*)(hpa + vmm_buf_pos), TDX_RANGE_RW);
    }

    vmm_buf_ptr[i] = msg_buf_ptr[i];
    vmm_buf_pos++;
  }
  ...
}

static void print_to_buffer(debug_control_t* p_ctl, char* print_buf, uint32_t
print_len) {
  ...
  debug_message_t* target_debug_message =
&p_ctl->trace_buffer[p_ctl->buffer_writer_pos];

  if (p_ctl->print_target == TARGET_EXTERNAL_BUFFER) {
    dump_message_to_vmm_memory(print_buf, (uint64_t)target_debug_message,
print_len, true);
  ...
}
```

Subsequent calls to logging functions, such as `tdx_print`, invoke `print_to_buffer`. In `print_to_buffer`, the address `target_debug_message` is calculated as `&p_ctl->trace_buffer[p_ctl->buffer_writer_pos]`. `target_debug_message` is then passed as the `hpa` argument to `dump_message_to_vmm_memory`. The `dump_message_to_vmm_memory` function calls `map_pa` to map this `hpa`. The `map_pa` function in `keyhole_manager.c` does not perform PAMT checks to prevent mapping of Intel TDX module private memory, if the HPA is within a TDMR.

As the host VMM can choose `trace_buffer` HPA and control `p_ctl->buffer_writer_pos`, by triggering a controlled number of log messages, it could point this to sensitive TDX module data (e.g., a TDCS page, TDR page, or other private data within a TDMR). The `dump_message_to_vmm_memory` function will then write up to `MAX_PRINT_LENGTH`, which is 256 bytes, of log data into this area.

**Remediation:** This code is only included in debug builds of the Intel TDX module and never in production binaries. Intel has opened an internal ticket and is tracking it as a low priority issue.

## Bug 11: Multiple Spectre Gadgets Enable for Out-of-Bounds Read but Are Unextractable

**Prerequisites**: Not Applicable.

Several subroutines accept user inputs that can result in accessing OOB data during speculative execution. However, since we cannot formulate a viable exploit for these gadgets, they are not classified as vulnerabilities, though Intel will apply DiD mitigations for them.

The subroutine `md_read_element` is used by various APIs to read the state of a TD. This subroutine indirectly calls `md_find_entry_idx` which uses the following code pattern to check that the user input `field_id.field_code` is within a valid range.

```
if ((field_id.class_code == lookup_table[i].field_id.class_code) &&
    ((uint64_t)field_id.field_code >= first_element_id_in_range) &&
    ((uint64_t)field_id.field_code < last_element_id_in_range)) {
  break;
}
```

Speculative execution of the above code results in accessing OOB data, which in theory is returned to the VMM or TD (depending if `md_read_element` is reached via a TDH or TDG interface).

Unlike Vulnerability 3, 4, and 5 the OOB data in this case is not processed by any succeeding code inside the Intel TDX module, hence it is not encoded to a microarchitectural state, like the cache, after the execution of the API. This is also not possible past the context switching because `SEAMRET`, `VMLAUNCH`, and `VMRESUME` block speculative execution, making the gadgets unexploitable.

Similarly, the following code checks if the index to the performance counter table `pmc_index` is within a valid range when emulating `RDMSR` execution for a TD. `rdmsr_ia32_perfevtsel` uses `pmc_index` to index a table and retrieve values to return to the TD via `RDX` and `RAX`.

```
static uint32_t get_pmc_index_given_ia32_perfevtsel_index(const uint32_t msr_addr)
{
  uint32_t invalid_idx = (uint32_t)INVALID_PERFMON_MSR_INDEX;

  // Legacy range
  if ((msr_addr >= IA32_PERFEVTSEL0_MSR_ADDR) && (msr_addr <
IA32_PERFEVTSEL0_MSR_ADDR + NUM_PMC)) {
```

```
    return msr_addr - IA32_PERFEVTSEL0_MSR_ADDR;
  }
  ...
}

_STATIC_INLINE_ td_msr_access_status_t rdmsr_ia32_perfevtsel(tdcs_t *tdcs_p,
tdvps_t *tdvps_p, uint32_t pmc_index, uint32_t msr_addr) {
  ...
  ia32_perfevtsel_t perfevtsel_value = { .raw =
tdvps_p->guest_msr_state.ia32_pmc_gp_cfg_ax[pmc_index] };
  perfevtsel_value.forbidden = 0;

  rdmsr_set_value_in_tdvps(tdvps_p, perfevtsel_value.raw);

  return TD_MSR_ACCESS_SUCCESS;
}
```

Lastly, when emulating `CPUID` execution leaf 0x4 uses a similar pattern to ensure the requested subleaf is within a valid range.

```
case 0x4:
  if ((td_ctls.reduce_ve) && (subleaf < NUM_CPUID4_NATIVE)) {
    ...
    return_values =
vp_ctx->tdcs->executions_ctl2_fields.cpuid4_native_values[subleaf];
    ...
  }
```

Although speculative execution of the above code sequences results in accessing OOB data, the speculation stops when `VMLAUNCH` or `VMRESUME` are executed, preventing a TD from being able to extract the data.

**Remediation**: Intel plans to review each case independently to understand the risk versus performance impact of adding `LFENCE`. It is expected that most if not all will be addressed as DiD. The `RDMSR` and `CPUID` emulation speculation primitives are only present in the early release of an updated Intel TDX 1.5 provided by Intel for this review.

# Review Methodologies

In this section, we discuss some of the techniques and tools we used for evaluation.

## TDXplore Toolkit

To support analysis of the Intel TDX Module the TDX Explore Toolkit (TDXplore) was developed to provide generic access to functionality normally reserved to ring-0 host VMM software. TDXplore is composed of three main components: a Linux kernel module, C/Python library, and a set of Python scripts that provide access to Intel TDX Module interfaces.

The Linux kernel module was developed to expose privileged functionality to userspace. This includes the ability to map/unmap physical memory, read/write kernel memory, and execute privileged instructions (e.g., `RDMSR`, `WRMSR`, `VMPTRLD`, `VMCLEAR`, `VMXON`, `VMREAD`, `VMWRITE`, `SEAMCALL`, and `TDCALL`). Most functionality is accessed via a set of ioctl calls with access to physical memory being accessible through `mmap` and `munmap`. The C/Python libraries simply wrap the Linux kernel module to provide a more user-friendly layer of abstraction.

Most of the Python scripts provide direct support for a specific interface in the Intel TDX Module. When combined or chained together they can be used to perform larger activities including:

- **TD and VP Management:** Initialize or create a TD and its VPs, add private memory pages, and configure the TD/VPs.
- **TD Migration:** Create a migration stream, bind a service TD to a target TD, pause a TD, and abort a migration.
- **Metadata and State Control:** Read and write metadata from the host VMM, service TD, and guest TD context.
- **Metadata Manipulation:** create, decrypt, parse, modify, and encrypt migration bundles with a provided MSK.

The table below shows the implemented scripts and provides a brief explanation.

| Names | Description |
|---|---|
| `mig_bundle_encrypt.py` `mig_bundle_decrypt.py` `mig_bundle_parse.py` `mig_bundle_edit.py` | Used to encrypt, decrypt, and interact with migration bundles. The MSK is provided to the encrypt and decrypt scripts as a parameter and the MBMD data is checked on decrypt and updated on encrypt. The parse and edit scripts work with decrypted immutable, td, and vp migration bundles. |
| `qemu_break.py`<br>`qemu_resume.py`<br>`qemu_stop.py` | The break script is used to watch for and suspend a QEMU process before it executes a specific `ioctl`. This is primarily used to pause execution before `KVM_TDX_FINALIZE_VM` is called to support binding a migTD using `tdh_servtd_bind.py`. The resume and stop scripts are simple wrappers for `kill` using `SIGCONT` and `SIGSTOP`. |
| `tdg_md_rd.py`<br>`tdg_md_wr.py`<br>`tdg_servtd_rd.py`<br>`tdg_servtd_wr.py` | Used to interact with TD metadata from within a TD. The `md` variant allows a TD to read and write its own metadata. The `servtd` variant supports reading and writing metadata of another TD from a service TD. |
| `tdh_export_abort.py`<br>`tdh_export_pause.py`<br>`tdh_export_state_immutable.py`<br>`tdh_export_state_td.py`<br>`tdh_export_state_vp.py` | Provides the ability to export migration bundles associated with a TD. The pause script moves a migration from the `OP_STATE_LIVE_EXPORT` to `OP_STATE_PAUSED_EXPORT` which is required to access TD and VP non-memory state. |
| `tdh_import_state_immutable.py`<br>`tdh_import_state_td.py`<br>`tdh_import_state_vp.py` | Provides the ability to import migration bundles to a previously created TD template. |
| `tdh_md_rd.py`<br>`tdh_md_wr.py` | Used to interact with TD metadata from the host VMM. |
| `global_sys_metadata.py`<br>`tdr_tdcs_metadata.py`<br>`tdvmcs_metadata.py`<br>`tdvps_metadata.py` | Metadata lookup lists are ported from `include/auto_gen_1_5` in the Intel TDX module source code. These are used by various other scripts parse and display entries (e.g., `tdh_md_rd.py`, `mig_bundle_parse.py`, and `tdg_servtd_rd.py`). |
| `tdh_servtd_bind.py`<br>`tdh_mig_stream_create.py` | Used to associate a service TD with a target TD and create migration stream contexts. |
| `tdh_mng_create.py`<br>`tdh_mng_key_config.py`<br>`tdh_mng_addcx.py`<br>`tdh_mng_init.py`<br>`tdh_vp_addcx.py`<br>`tdh_vp_create.py` | Various scripts to create and configure a TD and the associated VPs. The `tdh_mng_addcx.py` and `tdh_vp_addcx.py` scripts add or assign pages of memory to be used by either the TD or its VPs. |

| `tdxtend.py`<br>`tdxamine.py` | The `tdxtend.py` script is a wrapper for the gateway script providing access to Intel TDX specific data structures, interfaces, and error codes. It's also used to inspect processes that use KVM to extract TDR, TDCS, and TD VP Root (TDVPR) HPAs.<br><br>The `tdxamine.py` script is used to store and lookup TD associated HPAs by name or PID to be used with other scripts. It also stores shared state created by other scripts. |
| --- | --- |

The following example demonstrates how to add a TD to the `tdxamine` state. The `add_td_by_pid` sub-command uses the specified QEMU PID to lookup HPAs associating it with the provided name when added. The `print_state` subcommand lists the known TDs and details that were populated during its addition or when running other scripts. The `tdh_import_state_immutable` script then loads an immutable migration bundle using a TDR HPA which was looked up using the `print_tdr_pa_from_name` subcommand of `tdxamine`.

```
python tdxamine.py add_td_by_pid `pgrep -f -o qemu` mig_td
...
python tdh_servtd_bind.py $(python tdxamine.py print_tdr_pa_from_name dst_td)
$(python tdxamine.py print_tdr_pa_from_name mig_td)

python tdxamine.py print_state
td: name - mig_td, tdr_ka - 0xffffffffffffffff, tdr_pa - 0x60bde35000 hkid -
0xffffffffffffffff
  tdvpr 0: pa -  0x60bbe9a000
  ...
  tdvpr 15: pa -  0x60b917f000
  bind 0: handle - 0xe3029dce5af581d9
  bind 0: uuid - 1f1308f0811d80bb-7c11ab60de61cabe-86f40fb10759d15a-367130596bc3cd9
td: name - dst_td, tdr_ka - 0xff2601d300103000, tdr_pa - 0x103000 hkid - 0x10
  tdcs 0: pa -  0x102000    ka - 0xff2601d300102000
  ...
  migsc 0: ka - 0xff2601d300107000, pa -  0x107000
...

python tdh_import_state_immutable.py $(python tdxamine.py print_tdr_pa_from_name
dst_td) immutable.mbmd immutable.data
```

Toolkit development used a bare metal C3 GCP instance with Ubuntu 24.10 following the [Intel Trust Domain Extensions (TDX) on Ubuntu](#) setup instructions. With this setup KVM and QEMU are used to create, manage, and destroy TDs. This reduces complexity of the TDXplore framework while providing access to complex TD environments.

One issue with this approach is that most Intel TDX Module API's expect either a TDR or TDVPR HPA and KVM doesn't provide this in a generic way. To overcome this the toolkit adds an ioctl to lookup the kernel `struct file` pointer for a provided FD and PID. The structures holding these HPAs are `struct kvm` and `struct kvm_vcpu` and pointers for these are stored in the `private_data` field of their associated `struct file`.

Locating the correct FD for each can be done using `/proc/$PID/fd`. The FD with the symbolic link to `anon_inode:kvm-vm` contains `struct kvm` and `anon_inode:kvm-vcpu:{INDEX}` holds `struct kvm_vcpu`.

TDXplore uses specific offsets to locate the TDR and TDVPR HPAs within these structures with support for Ubuntu 24.10 (`tag: Ubuntu-intel-6.11.0-1008.8`) and Ubuntu 24.04 (`tag: Ubuntu-intel-6.8.0-1022.29`).

*Note: the TDXplore Linux kernel module is not appropriate for production environments because it fundamentally undermines userspace isolation from privileged operations.*

This toolkit empowered the team to conduct security research with a production version of the Intel TDX Module on hardware. Support for configuring a TD into different states, interacting with provided interfaces, and development of proof-of-concept exploits as bugs were identified was extremely important to understanding how the Intel TDX Module operates.

## LLM Bug Hunting

The Intel TDX 1.5 code review presented a good opportunity to assess the capabilities of LLMs (specifically Google Gemini) for identifying vulnerabilities in a complex and real world codebase. In particular, it had:

- **Limited dependencies:** The Intel TDX firmware code essentially only depends on the [Intel IPPS crypto library](#) but otherwise encompasses an entire operating system and business logic. This reduces the number of assumptions the LLM needs to make about libraries or syscalls outside of the context window.
- **Clean API organization:** Each API is divided into its own source file and furthermore VMM/TD source files are separated. This made submitting subsections of the Intel TDX code to the LLM trivial.

Next, we discuss how we used LLMs for this review to identify Spectre gadgets and memory safety vulnerabilities, and the current limitations of our approach.

## Spectre Gadgets

Our goal here was to identify code patterns for Spectre v1 gadgets inside the Intel TDX Module, as one of the limitations of current processors is that there is no hardware remedy for mitigating this class of attacks. As a result, critical software systems such as the Linux kernel and Intel TDX Module firmware relies on an ad-hoc approach to only apply fixes (e.g., using a serialization instruction like `LFENCE`) to code gadgets that are identified as exploitable. Once an attacker finds an exploitable gadget inside the Intel TDX Module firmware, it can potentially leak the entirety of private memory for the module and guests, while mapped into the module's linear address space.

We used two different Gemini models in a two-step approach to identify potential Spectre code gadgets. For this, we first describe the conditions for a valid Spectre gadget alongside the source code and specification as context to Gemini in thinking mode (slow but more capable) to identify potential gadgets, then, we use a faster but less capable mode like Gemini flash to summarize the findings and highlight key aspect of the findings, as follow:

| First Query |
| --- |
| You are an expert vulnerability researcher who knows Spectre vulnerabilities and their exploitations very well. Review the API function {api} its subroutines and identify potential Spectre v1 gadgets that allow accessing out-of-bounds data.<br><br>Such Spectre v1 gadgets have three components: a) User input: An input that is controlled by user, b) Branch condition that checks the user input c) secret-dependent memory access or branch depending on the out-of-bounds access.<br><br>Check if it has already been mitigated with an LENCE.<br><br>Summarize the findings and each identified vulnerability as follow:<br> - User Input:<br> - Branch Condition:<br> - Secret-Dependent Memory Access:<br> - Step-by-step analysis of the vulnerability:<br> - How Much information can be leaked:<br> - If it has already been mitigated: |

**Second Query**

Generate a summary of each discovered vulnerability with the following markdown format:
>    ## API Name: Vulnerability Name
>    ### User Input:
>    ### Branch Condition:
>    ### Secret-Dependent Memory Access:
>    ### Step-by-step analysis of the vulnerability:
>    ### How Much information can be leaked:
>    ### If it has already been mitigated:

We used the above queries for each API of interest including the host interface (`tdh_*`), the guest interface (`tdg_*`), and the guest exit handlers (`td_*_exit`), for a total of 97 APIs. In this setup, each API took about 3 minutes to analyze, a total of 5 hours to execute the combined queries.

As mentioned above, the clean API organization of each API having its own source file made it easier for us to use Gemini for this goal. However due the context window's limit (1 million tokens), we could not provide the entire source code and the ABI specification for Intel TDX module firmware as context. Instead, we used some custom scripts to slice each API and its dependencies, broke the ABI specification for each API to a separate PDF file, and only provided relevant information for each API to it.

Our initial strawman approach resulted in about 200 reports. After going through these reports, we identified the following false positives:

- **Duplicates:** Almost half of the reports were duplicates, and easy to filter out (i.e.., identifying the same code pattern as vulnerable in the dependency of every API).
- **Loop bounds:** Tens of cases of loop bounds being identified as potential Spectre gadgets. While these cases may theoretically be Spectre gadgets, they are hard to exploit in practice.
- **Public memory:** Several gadgets would result in OOB access to memory that is already accessible to the host VMM or guest TD, which shows Gemini did not understand the memory mappings and threat model well.
- **Union bounds:** Several gadgets were related to bounds that are implicitly enforced via C unions, hence the software bounds check even if bypassed speculatively does not result in OOB access.

After filtering out the above cases, we identified 16 code gadgets that can potentially leak private memory. This process took about 3 days, however, some of it can further be automated by querying the LLM to filter / group findings together. Among these 16 unique code gadgets, we identified:

- **9 Already fixed:** Mitigated with an `LFENCE` highlighting that they were indeed valid, but already mitigated.
- **5 New gadgets:** Acknowledged to impact security by Intel, and **three** CVE were assigned based on the root case.
- **2 DiD:** Only exploitable on a hardware architecture that would not serialize `SEAMRET`, `VMLAUNCH`, `VMRESUME` highlighting an additional lack of context surrounding hardware architecture limiting Gemini's analysis.

## Memory Corruption

We also spent time using Gemini to search for memory corruption and logic bugs in the C source code. For these problems, the LLM must correctly identify pathways for user input to reach unsafe operations while also meeting all necessary conditions that the code previously checked against. The LLM may also need to correctly reason about multi-threading race conditions and lock semantics depending on the bug.

Fundamentally, we started by asking Gemini to find memory corruption bugs and iterated on the prompts as we found deficiencies in the results. While there is some nuance to identifying the best prompting and LLM settings, the fundamental bottleneck in the process is triaging the bug reports and verifying correctness. We found that once the settings are reasonably tweaked, Gemini will generate bug reports that appear to be accurate and often if inaccuracies are found that they are similar to mistakes human reviewers would also make – this makes triaging time-consuming. We should also note that Gemini sometimes still makes basic mistakes like skipping over lines that do bounds checks right before an array access.

While the TDXplore toolkit that we developed could facilitate automated PoC generation, we did not investigate this during the review since the LLM work preceded most of the test framework development. Even then, setting up the Intel TDX state machine correctly and generating the precise inputs to trigger a vulnerability are relatively complex compared to other targets (e.g., a usermode binary or remote server) and we expect current LLMs to struggle here without extensive assistance.

Instead of depending on the LLM to generate PoCs for its own bug reports, we instead asked the LLM to do an initial triage pass which filters down the reports before a human does the final analysis. One intuition we had is that for a given LLM conversation, the LLM seems resistant to changing its mind about things; e.g., that a bug exists or a bounds-check that does exist was skipped over. By asking the same LLM to validate the previous LLM-generated response, this gives each request a higher likelihood of discovering mistakes compared to just asking the initial LLM session to self-check.

We essentially split the bug hunting task into this pipeline:

- Identify all APIs of interest (or we can manually generate/save this list).
- For each API of interest, ask the LLM to search for bugs N times.

○ Deduplicate results if N > 1.
- For each bug report, ask the LLM to validate the report M times.
  ○ Summarize the consensus arguments for true vs false positive.

## Limitations

Despite the presence of some inaccurate reports, Gemini helped narrow down analysis in several cases, and was particularly effective in identifying Spectre code gadgets. We identified the following limitations to help inform future use of LLMs for code review:

- **Unusual threat model:** In Intel TDX, almost all code is considered untrusted which leads to unusual circumstances such as APIs that can corrupt or leak VMM memory being non-interesting if initiated by the host. Additional prompting was used to help reduce false positives due to this confusion.
- **High level of out-of-context assumptions:** The Intel TDX Module code heavily depends on hardware functionality and makes many implicit assumptions based on this. While this hardware behavior is mostly defined in public specifications and likely incorporated in Gemini's training data, it's believed that not including it in the context window makes mistakes more likely.
- **Large code base:** All of the Intel TDX Module code encompasses around 2MiB which roughly translates to just over 1M tokens. Currently, Gemini 2.5Pro has a maximum input context length of 1M tokens so we've had to submit partial code to the LLM. Additionally, if one also wants to send specs with the code, further code reductions are required.
- **Ambiguous semantics:** There were several areas where the LLM consistently got confused due to unusual APIs or unclear intentions of the code authors. For example, the Intel TDX Module does not use a heap per-se but does have dynamically mapped memory using a "keyhole" system – the LLM would sometimes incorrectly identify use-after-free vulnerabilities that incorrectly assumed normal heap properties. Additionally, the Intel TDX Module has fairly complex multi-layer locks whose semantics are only partially defined in the code and specification.
- **Manual post-mortem analysis:** Currently, the Intel TDX Module toolchain is too complex with limited debugging capabilities, which makes it difficult to develop an end-to-end toolchain to let LLMs automatically evaluate findings and receive feedback (e.g., by triaging the bugs or trying to develop proof-of-concept exploits in a live system). As a result, manual expert analysis is needed to verify each finding.

## Frama-C

We used an off-the-shelf Frama-C Weakest Precondition (WP) plugin to do some basic code analysis. We compiled the Intel TDX Module 1.5 and generated the necessary symbol files so we can use Frama-C to identify potential memory safety violations. The compilation by itself identified several places where variables are assigned but never used, which is a weak code pattern. We also identified inconsistency across different APIs to initialize pointers, which one of them was also reflected in the [Vulnerability 4](#), which allowed for arbitrary values to be speculatively injected into the stack. The WP plugin flagged several memory safety cases, but manual review showed that those are all false positives, due to Frama-C failing to understand the union data structures and their bounds correctly.

## Code Difference

We did a limited differential analysis between the Intel TDX 1.0 and 1.5 source code to understand what changes were made and if there were any bug fixes which might have variants. One such difference that caught our attention was the change to `basic_memset` shown below.

```c
void basic_memset(uint64_t dst, uint64_t dst_bytes, uint8_t val, uint64_t nbytes) {
  tdx_sanity_check (dst_bytes >= nbytes, FATAL_ERROR_ID_176, 2);

  volatile uint64_t junk;

  _ASM_VOLATILE_ ("cld\n"
                  "rep; stosb;"
+                 :"=D"(junk) // marking that RDI is changing
                  :"c"(nbytes), "a"(val), "D"(dst)
                  :"memory", "cc");
}
```

A quick refresher on C inline assembly syntax and x86 instructions is probably helpful. The `REP; STOSB` statement is really the `STOSB` (store string byte) [instruction](#) with the `REP` prefix – this indicates that the operation should execute `RCX` times. `STOSB` copies data from the source pointer in `RSI` to the destination pointer in `RDI`. The `CLD` instruction clears the direction flag which causes the copy to auto-increment instead of auto-decrement.

The three lines following the x86 assembly are the output operand, input operands, and clobber list. The input/output operands indicate which registers are used as inputs and outputs. The clobber list indicates all other state that is somehow changed by the assembly code. In this case, the difference shows that `RDI` was added to the output operand list – this is accurate since `STOSB` updates `RDI` as the pointer increments or decrements. Without this hint, the compiler is free to optimize usage of `RDI` and could potentially assume it was unchanged and not reload the original value of `dst` into `RDI`.

Looking at this difference, we realized that this is an interesting bug pattern that we might have otherwise overlooked, which resulted in the discovery of [Bug 5](#). There are a handful of assembly helper functions in the Intel TDX Module and they all depend on accurate operand and clobber listings to ensure correctness. Going back to `basic_memset`, you may have noticed that another register is also changed by the assembly but not listed in the output: `RCX`. According to the `REP` prefix, `RCX` is decremented until reaching zero so therefore this register should be in the clobber list.

We used a combination of manual analysis and Gemini 2.5 Pro to analyze the remaining assembly helper functions for similar issues. The LLM approach was helpful since there are many implementations to check and the x86 instruction operations can be complex – this gave us a quick estimate of overall problem space.

Similar issues were discovered in these helpers:

- `ia32_rdrand,ia32_rdseed`: `RDRAND` modifies `CF` and `PUSHFQ`/`POPQ` interact with flags. `cc` should be clobbered.
- `_lock_read_128b`: `CMPXCHG16B` modifies flags (`ZF`). `cc` clobber is missing.
- `_lock_or_16b` (and related): `ORW`, `ANDB`, and `XORW` all modify flags (`OF`, `SF`, `ZF`, `PF`; `CF` cleared). `cc` clobber is missing for all these.
- `clear_xmms`: XMM registers `XMM0` through `XMM15` are modified and should be listed in the clobber list.
- `load_xmms_from_buffer`: XMM registers `XMM0` through `XMM15` are modified and should be listed in the clobber list.
- `clear_ymms`: YMM registers `YMM0` through `YMM15` are modified and should be listed in the clobber list.
- `load_ymms_from_buffer`: YMM registers `YMM0` through `YMM15` are modified and should be listed in the clobber list.
- `calculate_local_data`: `RDGSBASE` does not modify `RFLAGS`. `cc` clobber is unnecessary.
- `calculate_sysinfo_table`: `RDFSBASE` does not modify `RFLAGS`. `cc` clobber is unnecessary.

## Notebook LM for Cross Referencing

To aid in the review, [NotebookLM](#) was used as a centralized and searchable resource. All relevant Intel manuals, whitepapers, and source code were loaded, which was useful for researching key architectural aspects. The ability to quickly cross-reference source material and specifications helped validate output and locate relevant sections for further reading and reviewing. Notebook LM was not used to query for bugs in the codebase.

# Negative Results

In this section, we note some of the cases where we failed to identify a vulnerability despite a valid attack vector, showing that sufficient mitigations were in place:

## Impact of Live Migration on CPUID and MSR Instruction Behavior

Intel TDX Architecture impacts behavior of some instructions, virtualization of MSR registers and the virtualization of CPU features and how they are being exposed to the guest through `CPUID` instruction. The details are well documented in the Intel TDX Module Specification and the security implications of those behaviors are out of scope of this investigation but our investigation focused on whether the live migration of a CVM to a different host impacts the behavior of these instructions, CPUID values, or MSR virtualization.

Furthermore, the introduction of Virtualization Exception (#VE) Reduction allows a guest to change how the Intel TDX Module handles CPUID and MSRs, creating a complex state machine that must maintain consistent security guarantees across migrations.

In summary, The Intel TDX Module is designed to provide the following security and confidentiality guarantees after a CVM migration:

- **CPUID Virtualization:** The module ensures that `CPUID` leaves/subleaves configured during TD initialization remain consistent for the lifetime of the TD. Other fields are either fixed or reflect the native hardware capabilities, based on their security implications.
- **MSR Virtualization:** The module allows the VMM to configure some MSR values exposed to the guest and verifies the compatibility of these values during migration.
- **REDUCE_VE:** This feature allows the guest to relax certain restrictions, enabling a #VE handler to manage the virtualization of some `CPUID` leaves and MSRs.

We focused on `CPUID` fields that are calculated by the module or exposed natively from the host CPU, checking if hardware differences on a target host would alter their values post-migration.

Most of these fields relate to CPU capabilities that primarily affect workload performance. We found **no indication that changes in these values would impact the security or confidentiality** of the CVM. However, a small set of fields that can alter execution behavior requires more careful review below.

| Leaf (Register) | Description |
|---|---|
| 0x1 (EBX) | **Field Name:** Initial APIC ID<br>**Virtualization Type:** Calculated<br>**Virtualization Details:** TDVPS.VCPU_INDEX[7:0]<br>**Comments:** Initialized at TD Init time hence controlled by Intel TDX Module. |
| 0x1 (ECX) | **Field Name:** OSXSAVE<br>**Virtualization Type:** Calculated<br>**Virtualization Details:** CR4.OSXSAVE<br>**Comments:** Calculated through XFAM hence controlled by Intel TDX Module. |
| 0xD (EBX)<br>Sub-leaf 0x0 and 0x1 | **Field Name:** Max Bytes for Enabled Features<br>**Virtualization Type:** Calculated<br>**Virtualization Details:** Native<br>**Comments:** Calculated through XFAM hence controlled by Intel TDX Module. |
| 0x80000000 (EAX) | **Field Name:** Maxindex<br>**Virtualization Type:** Native<br>**Virtualization Details:** N/A<br>**Comments:** Highest calling parameter for CPUID. Would only impact CPUID enumeration code with no security risk. |
| 0x80000001 (EDX) | **Field Name:** SYSCALL/SYSRET in 64-bit Mode<br>**Virtualization Type:** Native<br>**Virtualization Details:** N/A<br>**Comments:** All CPU families supporting Intel TDX would support syscall hence this would be always 1. |
| 0x80000008 (EAX) | **Field Name:** Number of Linear Address Bits<br>**Virtualization Type:** Native<br>**Virtualization Details:** N/A<br>**Comments:** Configured by the host VMM's GPAW setting, virtualized by Intel TDX Module. |

We also reviewed the impact of `REDUCE_VE` on virtualization during live migration.

- With `REDUCE_VE`, some CPUID values that were previously managed by Intel TDX Module can now be emulated by the VMM. While this introduces a theoretical security risk, we **found no specific case where live migration created a vulnerability**.

- For MSRs, `REDUCE_VE` can change the result of an operation from a #VE to a General Protection (#GP). However, we **found no cases where this behavior would be different after live migration**.

## Ciphertext Side Channels Via Live Migration APIs

Another attack vector we considered was software-based [ciphertext side channels](#) that can bypass constant-time coding or potentially leak low-entropy values of registers. The `tdh_export_*` APIs could be vulnerable to such attacks if an attacker could export the same memory content, with the same key, and IV more than once.

Imagine the following common code pattern to mitigate recovery of a cryptographic key with a side channel. The code prevents secret dependent leakage due to `secret[i]` with help of an AND gate masking.

```
for(int i = 0; i < secret_len; i++) {
    int val_if_secret_true = (some_value + secret[i]) * 2;
    int mask = -(int)(secret[i] != 0);
    *target_page = (val_if_secret_true & mask);
}
```

If a ciphertext side-channel attack learns the secret by capturing two ciphertext for `target_page` before and after the memory store and if they differ, it learns the value of the `secret[i]`.

However, this attack is mitigated in these APIs because of the incrementing Initialization Vector (IV) counter. As an example, line 76 in `tdh_export_mem.c` shows `iv_counter++` with a comment stating its to prevent reuse.

```
// Increment the IV counter so we don't reuse a previous IV even if aborted
    migsc_p->iv_counter++;
```

As a result, two calculated ciphertext for a target memory page are expected to always have different values.

We also considered the fact that there is no bounds checking for this counter, which means at some point it could reset to a prior value. Imagine the following attack scenario:

1. Attacker blocks access to `target_page`

2. Attacker exports a `ciphertext_1` for `target_page`
3. Attacker lets the loop execute for one iteration (can be done by repeatedly block/unblock access to `target_page`)
4. Attacker executes `tdh_export_*` many times to increment the counter for $2^{64}$
5. Attacker exports a second `ciphertext_2` for `target_page` with the same IV and Key
6. If `ciphertext_1` and `ciphertext_2` differ, the attacker learns the `secret[i]` was set

However, there is a big challenge to conduct this attack, which is the computational complexity of incrementing the counter $2^{64}$ times. In `tdh_export_mem`, an attacker can increment this counter 512 times by every execution of this API (using `GPA_ENTRY_OP_NOP`).

Further, the Intel TDX module only allows a single thread to execute the migration stream, which limits an attacker's ability to scale the hypothesized attack.

```
// Lock the MIGSC link
if (!(migsc_lock(&tdcs_p->f_migsc_links[migs_i]))) {
  TDX_ERROR("Failed to lock tdcs_p->f_migsc_links[%u]\n", migs_i);
  return_val = api_error_with_operand_id(TDX_OPERAND_BUSY, OPERAND_ID_MIGSC);
  goto EXIT;
}
migsc_locked_flag = true;
```

A rough calculation without parallelization shows that this is not practical assuming it takes 10,000 cycles to increment the counter by 512 on each invocation of the API: 10,000 cycles * $2^{64}$ / (5 * $10^9$) / (60 seconds * 60 minutes * 24 hours * 365 days) / 512 increments = 2284.93 years.

## Exploiting Spectre Gadgets Past VMLAUNCH/VMRESUME

We mentioned in [Bug 11](#) that two of the Spectre gadgets we identified were not exploitable. We tried to develop a working exploit for one of these gadgets which would have allowed a guest TD to leak memory of the VMM via executing an `RDMSR` instruction which is emulated by the Intel TDX Module `RDMSR` exit handler and the results are output to the guest TD. This attack vector could have been easy to exploit considering that the output of the `RDMSR` is forwarded to the TD guest and that it can easily encode any forwarded value to cache accesses in the TD memory space.

However, our exploit code did not result in leaking OOB memory. After reporting the issue to Intel, they confirmed that speculation does not continue past `VMLAUNCH` or `VMRESUME` instructions, which confirms our observation.

# Recommendations

In this section, we discuss some of the recommendations for improving the security of the Intel TDX Module. Some recommendations focus on ways to reduce the attack surface of the Intel TDX Module while others suggest enablement of additional security technologies and known best practices.

## Memory Safety Mitigations via Segmented Linear Address Space

Defense-in-depth approaches are crucial to ensure mitigation against potentially undiscovered memory safety vulnerabilities. This is especially important because Intel TDX Module firmware is not developed in a safe programming language like Rust, and Spectre attacks can violate memory bounds checking.

The SEAM mode can read/write to all private memories that are mapped to the keyhole in the linear address. This means if there is a memory safety bug or Spectre gadget in the Intel TDX Module, the attacker can construct arbitrary memory addresses that target various physical memory (with HKIDs) and read arbitrary memory including TD private memory.

### Control Flow Integrity (CFI)

The Intel TDX module supports coarse-grain CFI based on Intel CET. The backward edge is protected by the CET shadow stack, and the forward edge by the landing pad `ENDBRANCH` instruction. These features make it difficult for an attacker to turn memory safety vulnerabilities into code execution inside the Intel TDX Module. The main caveat is that the landing pad instruction only provides coarse-grain CFI, hence an attacker may still be able to construct code reuse attacks.

One way to efficiently address this limitation is to apply [FineIBT](#), which combines the Intel CET landing pad instruction with cheap software label checks. As claimed by the authors, the performance overhead of this approach is less than 2%, and it has already been deployed in the Linux kernel.

### Guard Pages

Guard pages are a common defensive mitigation to prevent out-of-bounds memory primitives at the top-level memory region categorization. For example, threads typically have their own stack and these stacks are often mapped contiguously in a single large pool. A large stack overflow (or a small overflow at the top of the stack) can access the adjacent thread's stack. Guard pages can be inserted between these stacks to prevent this OOB access from accessing inter-thread memory.

These pages are typically implemented by separating each region by a single page width, then mapping in a page table entry that falls in this gap with permissions set such that the page is marked not present. No backing memory is required for the guard page and thus the only cost is a single PTE and a small portion of the virtual address space.

We recommend that Intel add similar guard pages between each thread's stack region. Currently these are separated by the CET shadow stack pages but these pages are still marked read-only. This prevents cross-thread linear OOB writes but does not prevent cross-thread linear OOB reads. We demonstrated this in Vulnerability 2 and Bug 1. The top-level memory regions (stack, global, code, keyhole) are already separated by unmapped memory and don't need additional guard pages.

## Software Fault Isolation (SFI)

A limitation of guard pages is that they do not protect against non-linear OOB violations—if an attacker can construct arbitrary addresses due to memory indexing overflow or Spectre gadgets, they can go over the guard pages.

One solution to this is to apply software fault isolation in addition to guard pages. In the Intel TDX Module, `map_pa` is essentially calculating a linear address. This linear address is never supposed to reach memory beyond an nGB region (there is no supergiant page mapping). If we can efficiently and reliably check that every memory access is within a nGB bounds, it should be possible to mitigate OOB memory safety violations (and Spectre gadgets) so that an attacker cannot overflow a linear memory address into an arbitrary region.

In x86_64, this bounds checking operation is almost free. Based on NaCL, one can simply encode all memory accesses as the following and implicitly enforce such bounds checking operation:

```
basereg + indexreg * scale + disp32

For example, in this pseudo-instruction:

add $0x00abcdef, %ecx
mov %eax, disp32(%RZP, %rcx, scale)
```

As a result, every memory access is limited to a 100G region, so that the memory that is mapped for other VPs or the TD private memory (e.g., during calling various APIs) is not accessible to a OOB vulnerability.

This potentially prevents OOB reads (including Spectre v1 gadgets) in a way that an attacker can only read what is supposed to be accessed by Intel TDX Module, for a lot of APIs, this is just metadata, not user data.

One caveat is that right now `mig_*_keys` are mapped inside the TDCS, which are highly-privileged credentials that, if leaked, lead to full compromise of TD security. But if SFI works above, you can also map those credentials to a separate linear space in a different 100 GB region, so they are not reachable directly.

SFI comes with execution overhead, although modern hardware extensions like Intel CET and MPK can be used to reduce the execution cost, more research and experimentation is required to assess if this is a practical solution.

## Reducing TCB via Attestable Global Feature Disablement

One of our learnings in this engagement is that the Intel TDX Module TCB is growing with every new feature (Live migration, TD Partitioning, TDX Connect, etc.), and this growth introduces a large attack surface for users. This is problematic in several scenarios:

- A feature may not be used by all customers, but vulnerabilities in the Intel TDX Module remain exploitable regardless.
- The Intel TDX Module sometimes receives functionality before other components are ready (e.g., MigTD, Host and Guest Kernel Support), so while a feature is not even usable, the attack surface is present.
- Completely addressing vulnerabilities in the Intel TDX Module can be time consuming due IPU and TCB Recovery cycles, leaving customers with no mitigation option until patches are applied.

We believe that the Intel TDX Module should have a set of global flags that are sticky, configured during initialization, and attestable. These flags could allow a host to enable only used features, enable only used interfaces, and lock TD attributes. This could limit the attack surface on a compromised host.

Examples include:

- Host VMM could disable the migration feature
- Host VMM disables loading of debuggable TDs via a global flag
- Host VMM disables loading of migratable TDs via a global flag

Currently, multiple Intel TDX Module features are already opted-in by the host VMM:

- TD Migration is opted-in by the TD's MIGRATABLE attribute.
- TD Partitioning is opted-in by configuring the number of L2 VMs to a non-0 value.
- TDX Connect is opted-in at the global level (`tdh_sys_config` and `tdh_sys_update` input flag) and per-TD.

- Perfmon events filtering is a VMM-configured feature.

We recommend also **supporting global enable/disable flags** to remove the possibility that TD security may be impacted by a feature, even if that feature is disabled for that specific TD, and the fact that a certain feature is globally disabled should be **attestable**.

Enablement or disablement could be performed during `tdh_sys_init` or earlier (e.g., PSEAMLDR, sticky MSR bits). This also potentially requires refactoring of the Intel TDX Module firmware to ensure that code related to a disabled feature is not reachable via any API / user input.

## Challenges with Reflecting Platform Configuration in Attestation Reports

Currently, the attestation report is largely static. If platform configuration is updated via runtime μcode patches, the changes are not reflected in the report until the TD Quoting SGX enclave is restarted. While TDQuotes generated after that enclave restart will reflect SVN changes in trusted components, a **critical gap** exists: there is no notification mechanism to inform the guest VM when to initiate a new TD Quote.

This gap forces an implicit trust in Cloud Service Providers (CSPs) to provide timely updates to TCB components. If the platform is patched for a known vulnerability, there is no way to guarantee the guest VM can verify that the system is running the safer version before a TCB component is compromised during that vulnerable period. Similarly, no notification mechanism exists to alert a guest VM of underlying TCB changes following a live migration.

While the Intel TDX Live Migration TCB is measured and included in the attestation report, it is crucial to understand that this TCB is a combination of the **Intel TDX LM MigTD binary** and the **Migration Policy** provided by the platform owner. Intel provides a reference MigTD implementation, but the current Intel TDX architecture does not enforce rules on the policy nor does it authenticate the MigTD binary.

These components are measured and reported, but it is fully the customer's responsibility to review this information and establish trust. CVM customers must carefully review the source of the MigTD binary and the security implications of the Migration Policy, verifying both as part of their attestation verification process.

Intel has recently published a proposal where the MigTD is not under the control of the CSP but is integrated into the TDX Module.

## Memory Safe Language and Formal Verification

The Intel TDX Module and NP/P-SEAMLDRs are written in C, which is not a memory safe language, and because of this considerable time was invested to review source code for memory safety issues. Formal verification of C code is problematic due to its memory model and direct interaction with hardware, which can result in undefined behavior. While tools exist, such as Frama-C for analysis and [CompCert](CompCert) for compilation, using a memory safe language, such as Rust, would have almost eliminated this entire class of vulnerabilities.

*Usage of unsafe in Rust would still need to be scrutinized but multiple vulnerabilities and bugs disclosed in this report would have not existed.*

Furthermore, multiple recommendations described above are included for the sole purpose of providing DiD against memory safety issues, which themselves increase complexity of the software and carry a non-zero impact to system performance.

Memory safe languages, such as Rust, are already being used for the development of Intel's MigTD and would be a great alternative for these codebases. Intel could rewrite the existing implementations, in a memory safe language, but allowing 3rd parties to develop their own TDX module provides the most flexibility. This would allow 3rd parties to select the language, build tools, desired features, and utilize analysis tools that meet their specific functional needs and security requirements.

# Disclosure Timeline

Following our investigation, we discovered and disclosed several security vulnerabilities to Intel. Intel promptly assigned CVEs but set a public disclosure timeline of **February 2026** for the following reasons:

1. **High-Risk Updates:** Updates to the Intel TDX Module are inherently disruptive to production environments due to the module's high privilege level. Therefore, simply patching the binary is insufficient. Each cloud provider will need several months to test, qualify, and safely roll out the fix to their infrastructure.
2. **Customer Attestation:** Any security fix that changes the TCB requires close coordination with customers. This collaboration is essential to give them time to update their attestation policies and prevent unexpected disruptions to their workloads.

Additionally, some identified items (less critical bug fixes and some security weaknesses) are not included in the February 2026 release but are expected to be addressed in subsequent releases.

As security researchers, we feel a responsibility to all Intel TDX users. We are adhering to **Intel's timeline** to ensure a proper mitigation is in place before the vulnerabilities are publicly detailed.

We have verified that **no Google CVM customers were exposed** to vulnerability 1. Following the principle of least privilege, Live Migration support has never been enabled in our production environment. This can be independently verified through the hardware-rooted

Intel TDX attestation report generated on-demand. The Google Cloud log also stores immutable copies of Hardware rooted attestation reports for CVM instances created in the past. Google has written a [blog post](#) to provide guidance on how attestation can be used to verify SVNs and attributes.

Furthermore, for vulnerability 2, 3, 4, and 5 we have verified that there has been **no evidence of active exploitation** of these vulnerabilities among Google CVM customers.

## Acknowledgments

The review team acknowledges the contributions of the following people for addressing technical questions and sharing security expertise.

We would like to extend our gratitude to the following Intel engineers: **Uri Bear**, **Dror Caspi**, **Stephen Haruna**, **Simon Johnson**, **Nagaraju Kodalapura**, **Alon Levi**, **Dhinesh Manoharan**, **Avishai Redelman**, **Bernie Reeber**, **Fahimeh Rezaei**, **Boaz Tamir**, and **Jonathan Valamehr**.

## Appendix A

The data structures used in by the Intel TDX Module for the `SEAMCALL` and `TDCALL` API located in `include/auto_gen/op_state_lookup`.c as `seamcall_state_lookup` and `tdcall_state_lookup`. Both tables are two-dimensional arrays of type `bool_t` so a couple of scripts were created to parse them into a more readable format.

This table shows the different `op_state` of a TD and the available `SEAMCALL` API. The notes provide the primary transition API and next state.

| Operation | Allowed SEAMCALLS | Notes |
|---|---|---|
| `OP_STATE_UNINITIALIZED` | `TDH_MNG_ADDCX_LEAF`<br>`TDH_VP_FLUSH_LEAF`<br>**`TDH_MNG_INIT_LEAF`**<br>`TDH_SERVTD_BIND_LEAF`<br>`TDH_SERVTD_PREBIND_LEAF`<br>**`TDH_IMPORT_STATE_IMMUTABLE`**<br>**`_LEAF`**<br>`TDH_MIG_STREAM_CREATE_LEAF` | **`TDH_MNG_INIT_LEAF` transitions to**<br>**`OP_STATE_INITIALIZED`**<br><br>**`TDH_IMPORT_STATE_IMMUTABLE_LEAF`**<br>**transitions to `OP_STATE_MEMORY_IMPORT`**<br><br>`TDH_IMPORT_STATE_IMMUTABLE_LEAF`<br>transitions to `OP_STATE_FAILED_IMPORT` |

| OP_STATE_INITIALIZED | TDH_MEM_PAGE_ADD_LEAF<br>TDH_MEM_SEPT_ADD_LEAF<br>TDH_VP_ADDCX_LEAF<br>TDH_MEM_PAGE_RELOCATE<br>TDH_MEM_PAGE_AUG_LEAF<br>TDH_MEM_RANGE_BLOCK_LEAF<br>TDH_VP_CREATE_LEAF<br>TDH_MNG_RD_LEAF<br>TDH_MEM_RD_LEAF<br>TDH_MNG_WR_LEAF<br>TDH_MEM_WR_LEAF<br>TDH_MEM_PAGE_DEMOTE_LEAF<br>TDH_MR_EXTEND_LEAF<br>**TDH_MR_FINALIZE_LEAF**<br>TDH_VP_FLUSH_LEAF<br>TDH_VP_INIT_LEAF<br>TDH_MEM_PAGE_PROMOTE_LEAF<br>TDH_MEM_SEPT_RD_LEAF<br>TDH_VP_RD_LEAF<br>TDH_MEM_PAGE_REMOVE_LEAF<br>TDH_MEM_SEPT_REMOVE_LEAF<br>TDH_MEM_TRACK_LEAF<br>TDH_MEM_RANGE_UNBLOCK_LEAF<br>TDH_VP_WR_LEAF<br>TDH_SERVTD_BIND_LEAF<br>TDH_SERVTD_PREBIND_LEAF<br>TDH_MIG_STREAM_CREATE_LEAF | **TDH_MR_FINALIZE_LEAF** transitions to<br>**OP_STATE_RUNNABLE** |
| OP_STATE_RUNNABLE | TDH_VP_ENTER_LEAF<br>TDH_MEM_SEPT_ADD_LEAF<br>TDH_MEM_PAGE_RELOCATE<br>TDH_MEM_PAGE_AUG_LEAF<br>TDH_MEM_RANGE_BLOCK_LEAF<br>TDH_MNG_RD_LEAF<br>TDH_MEM_RD_LEAF<br>TDH_MNG_WR_LEAF<br>TDH_MEM_WR_LEAF<br>TDH_MEM_PAGE_DEMOTE_LEAF<br>TDH_VP_FLUSH_LEAF<br>TDH_MEM_PAGE_PROMOTE_LEAF<br>TDH_MEM_SEPT_RD_LEAF<br>TDH_VP_RD_LEAF<br>TDH_MEM_PAGE_REMOVE_LEAF<br>TDH_MEM_SEPT_REMOVE_LEAF<br>TDH_MEM_TRACK_LEAF<br>TDH_MEM_RANGE_UNBLOCK_LEAF<br>TDH_VP_WR_LEAF<br>TDH_SERVTD_BIND_LEAF<br>TDH_EXPORT_RESTORE_LEAF<br>**TDH_EXPORT_STATE_IMMUTABLE_LEAF**<br>TDH_EXPORT_UNBLOCKW_LEAF<br>TDH_MIG_STREAM_CREATE_LEAF | **TDH_EXPORT_STATE_IMMUTABLE_LEAF transitions to** OP_STATE_LIVE_EXPORT |

| OP_STATE_LIVE_EXPORT | TDH_VP_ENTER_LEAF<br>TDH_MEM_SEPT_ADD_LEAF<br>TDH_MEM_PAGE_RELOCATE<br>TDH_MEM_PAGE_AUG_LEAF<br>TDH_MEM_RANGE_BLOCK_LEAF<br>TDH_MNG_RD_LEAF<br>TDH_MEM_RD_LEAF<br>TDH_MNG_WR_LEAF<br>TDH_MEM_WR_LEAF<br>TDH_MEM_PAGE_DEMOTE_LEAF<br>TDH_VP_FLUSH_LEAF<br>TDH_MEM_PAGE_PROMOTE_LEAF<br>TDH_MEM_SEPT_RD_LEAF<br>TDH_VP_RD_LEAF<br>TDH_MEM_PAGE_REMOVE_LEAF<br>TDH_MEM_SEPT_REMOVE_LEAF<br>TDH_MEM_TRACK_LEAF<br>TDH_MEM_RANGE_UNBLOCK_LEAF<br>TDH_VP_WR_LEAF<br>TDH_SERVTD_BIND_LEAF<br>**TDH_EXPORT_ABORT_LEAF**<br>TDH_EXPORT_BLOCKW_LEAF<br>TDH_EXPORT_MEM_LEAF<br>**TDH_EXPORT_PAUSE_LEAF**<br>**TDH_EXPORT_TRACK_LEAF**<br>TDH_EXPORT_UNBLOCKW_LEAF | **TDH_EXPORT_ABORT_LEAF** transitions to **OP_STATE_RUNNBALE**<br><br>**TDH_EXPORT_ABORT_LEAF** transitions to **OP_STATE_PAUSED_EXPORT**<br><br>**TDH_EXPORT_TRACK_LEAF** transitions to **OP_STATE_POST_EXPORT** |
| OP_STATE_PAUSED_EXPORT | TDH_MEM_PAGE_RELOCATE<br>TDH_MEM_RANGE_BLOCK_LEAF<br>TDH_MNG_RD_LEAF<br>TDH_MEM_RD_LEAF<br>TDH_MEM_PAGE_DEMOTE_LEAF<br>TDH_VP_FLUSH_LEAF<br>TDH_MEM_PAGE_PROMOTE_LEAF<br>TDH_MEM_SEPT_RD_LEAF<br>TDH_VP_RD_LEAF<br>TDH_MEM_PAGE_REMOVE_LEAF<br>TDH_MEM_SEPT_REMOVE_LEAF<br>TDH_MEM_TRACK_LEAF<br>TDH_MEM_RANGE_UNBLOCK_LEAF<br>**TDH_EXPORT_ABORT_LEAF**<br>TDH_EXPORT_MEM_LEAF<br>TDH_EXPORT_TRACK_LEAF<br>TDH_EXPORT_STATE_TD_LEAF<br>TDH_EXPORT_STATE_VP_LEAF<br>TDH_EXPORT_UNBLOCKW_LEAF | **TDH_EXPORT_ABORT_LEAF** transitions to **OP_STATE_RUNNBALE** |
| OP_STATE_POST_EXPORT | TDH_MEM_PAGE_RELOCATE<br>TDH_MEM_RANGE_BLOCK_LEAF<br>TDH_MNG_RD_LEAF<br>TDH_MEM_RD_LEAF<br>TDH_MEM_PAGE_DEMOTE_LEAF<br>TDH_VP_FLUSH_LEAF<br>TDH_MEM_PAGE_PROMOTE_LEAF<br>TDH_MEM_SEPT_RD_LEAF<br>TDH_VP_RD_LEAF<br>TDH_MEM_PAGE_REMOVE_LEAF<br>TDH_MEM_SEPT_REMOVE_LEAF<br>TDH_MEM_TRACK_LEAF<br>TDH_MEM_RANGE_UNBLOCK_LEAF<br>**TDH_EXPORT_ABORT_LEAF**<br>TDH_EXPORT_MEM_LEAF<br>TDH_EXPORT_UNBLOCKW_LEAF | **TDH_EXPORT_ABORT_LEAF transitions to OP_STATE_RUNNBALE** |
| OP_STATE_MEMORY_IMPORT | TDH_MEM_SEPT_ADD_LEAF<br>TDH_MNG_RD_LEAF<br>TDH_MEM_RD_LEAF<br>TDH_MNG_WR_LEAF<br>TDH_MEM_WR_LEAF<br>TDH_VP_FLUSH_LEAF<br>TDH_MEM_SEPT_RD_LEAF<br>TDH_VP_RD_LEAF<br>TDH_MEM_SEPT_REMOVE_LEAF<br>TDH_MEM_TRACK_LEAF<br>TDH_VP_WR_LEAF<br>TDH_IMPORT_ABORT_LEAF<br>**TDH_IMPORT_MEM_LEAF**<br>**TDH_IMPORT_TRACK_LEAF**<br>**TDH_IMPORT_STATE_TD_LEAF** | **TDH_IMPORT_STATE_TD_LEAF transitions to OP_STATE_STATE_IMPORT**<br><br>**TDH_IMPORT_TRACK_LEAF transitions to OP_STATE_POST_IMPORT**<br><br>TDH_IMPORT_MEM_LEAF transitions to OP_STATE_LIVE_IMPORT<br><br>TDH_IMPORT_ABORT_LEAF transitions to OP_STATE_FAILED_IMPORT<br><br>TDH_IMPORT_STATE_TD_LEAF transitions to OP_STATE_FAILED_IMPORT |
| OP_STATE_STATE_IMPORT | TDH_MEM_SEPT_ADD_LEAF<br>TDH_VP_ADDCX_LEAF<br>TDH_VP_CREATE_LEAF<br>TDH_MNG_RD_LEAF<br>TDH_MEM_RD_LEAF<br>TDH_MNG_WR_LEAF<br>TDH_MEM_WR_LEAF<br>TDH_VP_FLUSH_LEAF<br>TDH_MEM_SEPT_RD_LEAF<br>TDH_VP_RD_LEAF<br>TDH_MEM_SEPT_REMOVE_LEAF<br>TDH_MEM_TRACK_LEAF<br>TDH_VP_WR_LEAF<br>TDH_IMPORT_ABORT_LEAF<br>**TDH_IMPORT_MEM_LEAF**<br>**TDH_IMPORT_TRACK_LEAF**<br>**TDH_IMPORT_STATE_VP_LEAF** | **TDH_IMPORT_TRACK_LEAF transitions to OP_STATE_POST_IMPORT**<br><br>TDH_IMPORT_ABORT_LEAF transitions to OP_STATE_FAILED_IMPORT<br><br>TDH_IMPORT_MEM_LEAF transitions to OP_STATE_LIVE_IMPORT<br><br>TDH_IMPORT_STATE_VP_LEAF transitions to OP_STATE_FAILED_IMPORT |
| OP_STATE_POST_IMPORT | TDH_MEM_SEPT_ADD_LEAF<br>TDH_MEM_PAGE_RELOCATE<br>TDH_MEM_RANGE_BLOCK_LEAF<br>TDH_MNG_RD_LEAF<br>TDH_MEM_RD_LEAF<br>TDH_MNG_WR_LEAF<br>TDH_MEM_WR_LEAF<br>TDH_MEM_PAGE_DEMOTE_LEAF<br>TDH_VP_FLUSH_LEAF<br>TDH_MEM_PAGE_PROMOTE_LEAF<br>TDH_MEM_SEPT_RD_LEAF<br>TDH_VP_RD_LEAF<br>TDH_MEM_PAGE_REMOVE_LEAF<br>TDH_MEM_SEPT_REMOVE_LEAF<br>TDH_MEM_TRACK_LEAF<br>TDH_MEM_RANGE_UNBLOCK_LEAF<br>TDH_VP_WR_LEAF<br>TDH_SERVTD_BIND_LEAF<br>**TDH_IMPORT_ABORT_LEAF**<br>**TDH_IMPORT_END_LEAF**<br>**TDH_IMPORT_COMMIT_LEAF**<br>**TDH_IMPORT_MEM_LEAF** | **TDH_IMPORT_END_LEAF transitions to OP_STATE_RUNNBALE**<br><br>**TDH_IMPORT_COMMIT_LEAF transitions to OP_STATE_LIVE_IMPORT**<br><br>TDH_IMPORT_MEM_LEAF transitions to OP_STATE_FAILED_IMPORT<br><br>TDH_IMPORT_ABORT_LEAF transitions to OP_STATE_FAILED_IMPORT |
| OP_STATE_LIVE_IMPORT | TDH_VP_ENTER_LEAF<br>TDH_MEM_SEPT_ADD_LEAF<br>TDH_MEM_PAGE_RELOCATE<br>TDH_MEM_PAGE_AUG_LEAF<br>TDH_MEM_RANGE_BLOCK_LEAF<br>TDH_MNG_RD_LEAF<br>TDH_MEM_RD_LEAF<br>TDH_MNG_WR_LEAF<br>TDH_MEM_WR_LEAF<br>TDH_MEM_PAGE_DEMOTE_LEAF<br>TDH_VP_FLUSH_LEAF<br>TDH_MEM_PAGE_PROMOTE_LEAF<br>TDH_MEM_SEPT_RD_LEAF<br>TDH_VP_RD_LEAF<br>TDH_MEM_PAGE_REMOVE_LEAF<br>TDH_MEM_SEPT_REMOVE_LEAF<br>TDH_MEM_TRACK_LEAF<br>TDH_MEM_RANGE_UNBLOCK_LEAF<br>TDH_VP_WR_LEAF<br>TDH_SERVTD_BIND_LEAF<br>TDH_EXPORT_STATE_IMMUTABLE<br>_LEAF<br>**TDH_IMPORT_END_LEAF**<br>**TDH_IMPORT_MEM_LEAF** | **TDH_IMPORT_END_LEAF transitions to OP_STATE_RUNNBALE**<br><br>**TDH_EXPORT_STATE_IMMUTABLE_LEAF transitions to OP_STATE_LIVE_EXPORT**<br><br>**TDH_IMPORT_MEM_LEAF transitions to OP_STATE_LIVE_IMPORT** |
| OP_STATE_FAILED_IMPORT | TDH_MNG_RD_LEAF<br>TDH_MEM_RD_LEAF<br>TDH_MNG_WR_LEAF<br>TDH_MEM_WR_LEAF<br>TDH_VP_FLUSH_LEAF<br>TDH_MEM_SEPT_RD_LEAF<br>TDH_VP_RD_LEAF<br>TDH_MEM_PAGE_REMOVE_LEAF<br>TDH_MEM_SEPT_REMOVE_LEAF<br>TDH_VP_WR_LEAF<br>**TDH_IMPORT_ABORT_LEAF** | TDH_IMPORT_ABORT_LEAF transitions to OP_STATE_FAILED_IMPORT |

Another table was created for the `TDCALL` API but not included here due to its simplicity. It showed that all states allowed both `TDG_SERVTD_RD_LEAF`, and `TDG_SERV_TD_WR_LEAF` except `OP_STATE_PAUSED_EXPORT` and `OP_STATE_POST_EXPORT` which only allowed `TDG_SERVTD_RD_LEAF`.

# Appendix B

Example source code showing how excluding `RCX` from the clobber list can introduce bugs depending on usage and compiler optimization.

```c
// gcc version 14.2.0
// gcc -O2 clobber.c
// gcc -O1 clobber.c
// gcc -Os clobber.c

// clang version 19.1.7
// clang -O1 clobber.c
// clang -O2 clobber.c
// clang -Os clobber.c

#define _GNU_SOURCE
#include <string.h>
#include <stdint.h>
#include <stdio.h>
#include <stdlib.h>
#include <signal.h>

// #define INFINITE_LOOP_TEST
// #define PAGE_FAULT_TEST

// #define RCX_IN_INPUT_ONLY
// #define RCX_IN_INPUT_AND_OUTPUT

#ifdef RCX_IN_INPUT_ONLY
static inline void tdx_memcpy(void *dst, uint64_t dst_bytes, void *src, uint64_t
nbytes) {
    volatile uint64_t junk_a, junk_b;

    asm volatile("rep; movsb;"
                 : "=S"(junk_a), "=D"(junk_b)
                 : "c"(nbytes), "S"(src), "D"(dst)
                 : "memory");
}
#endif // RCX_IN_INPUT_ONLY

#ifdef RCX_IN_INPUT_AND_OUTPUT
```

```c
static inline void tdx_memcpy(void *dst, uint64_t dst_bytes, void *src, uint64_t
nbytes) {
    volatile uint64_t junk_a, junk_b, junk_c;

    asm volatile("rep; movsb;"
                 : "=S"(junk_a), "=D"(junk_b), "=c"(junk_c)
                 : "c"(nbytes), "S"(src), "D"(dst)
                 : "memory");
}
#endif // RCX_IN_INPUT_AND_OUTPUT

#define ARRAY_SIZE 64
uint8_t dst[ARRAY_SIZE] = {42};
uint8_t src[ARRAY_SIZE] = {73};

void sigsegv_handler(int signum, siginfo_t *si, void *context) {

    ucontext_t *uc = (ucontext_t *)context;

    printf("[-] test failed: sigsegv rip: 0x%llx, rcx: 0x%llx\n",
uc->uc_mcontext.gregs[REG_RIP], uc->uc_mcontext.gregs[REG_RCX]);
    exit(-1);
}

int main(int argc, char *argv[]) {

    struct sigaction sa;
    sa.sa_flags = SA_SIGINFO;
    sa.sa_sigaction = &sigsegv_handler;
    sigaction(SIGSEGV, &sa, NULL);

#ifdef PAGE_FAULT_TEST
    int count = 64;

    for (int i = 0; i < count; i++) {
        tdx_memcpy(dst, sizeof(dst), src, count);
        count--;
    }
#endif // PAGE_FAULT_TEST

#ifdef INFINITE_LOOP_TEST
    for (int i = 0; i < ARRAY_SIZE; i++) {
        tdx_memcpy(dst, sizeof(dst), src, i);
    }
#endif // INFINITE_LOOP_TEST

    printf("[+] test passed\n");
```

```
    return 0;
}
```